\documentclass[prd,aps,nofootinbib,floatfix,11pt]{revtex4}

\usepackage{amsmath,graphicx,epsfig,amssymb,dsfont,mathtools,}
\usepackage[usenames]{color}
\usepackage{ulem} 
\usepackage{bigstrut}
\usepackage{slashed}
\usepackage{multirow}
\usepackage{makecell}

\allowdisplaybreaks

\begin{document}
\title{Weak Decays of Doubly Heavy Baryons: SU(3) Analysis  }
\author{Wei Wang,  Zhi-Peng Xing~\footnote{Email:zpxing@sjtu.edu.cn}, and Ji Xu~\footnote{Email:xuji1991@sjtu.edu.cn}}
\affiliation{ INPAC, Shanghai Key Laboratory for Particle Physics and Cosmology, School of Physics and Astronomy, Shanghai Jiao Tong University, Shanghai  200240,   China }

\begin{abstract}
Motivated  by the recent LHCb  observation of doubly-charmed baryon $\Xi_{cc}^{++}$  in the   $\Lambda_c^+ K^-\pi^+\pi^+$ final state,  we analyze    the weak decays of doubly heavy baryons
$\Xi_{cc}$,  $\Omega_{cc}$, $\Xi_{bc}$, $\Omega_{bc}$, $\Xi_{bb}$ and $\Omega_{bb}$ under   the flavor SU(3) symmetry.  Decay amplitudes for various semileptonic and nonleptonic decays are parametrized in terms of a few  SU(3) irreducible amplitudes.  We find a number of relations or sum rules between   decay widths and CP asymmetries, which   can be examined in future measurements at experimental facilities like LHC, Belle II and CEPC. Moreover once a few decay branching fractions are measured in future,  some of these relations may provide hints for exploration  of new decay modes.
\end{abstract}

\maketitle

\section{Introduction}

The existence of doubly heavy baryons is predicted in the quark model, but   the experimental  search for doubly heavy baryons has been a while~\cite{Mattson:2002vu,Ocherashvili:2004hi,Kato:2013ynr,Aaij:2013voa,Aubert:2006qw,Ratti:2003ez}.
Recently, the LHCb collaboration  has observed the doubly charmed baryon $\Xi_{cc}^{++}$ with  the mass   given as~\cite{1707.01621}
\begin{eqnarray}
m_{\Xi_{cc}^{++}} = (3621.40\pm 0.72\pm 0.27\pm 0.14) {\rm MeV}. \label{eq:LHCb_measurement}
\end{eqnarray}
Without no doubt,  this observation will  make a great impact on the hadron spectroscopy and   it  will also   trigger   more  interests in this research field~\cite{Chen:2017sbg,Meng:2017udf}. Moreover
after this  observation, we also anticipate  more  experimental investigations of decays of doubly heavy baryons.   Thus   theoretical  studies on   weak decays of doubly heavy baryons  will be  of great importance and are  forcefully requested~\cite{Guo:1998yj,SanchisLozano:1994vh,Faessler:2001mr,Ebert:2004ck,Albertus:2006wb,Hernandez:2007qv,Flynn:2007qt,Albertus:2009ww,Faessler:2009xn,Li:2017ndo,Yu:2017zst,Wang:2017mqp}.

QCD as the fundamental theory for strong interactions shows two distinct facets. At high energy,  the interaction strength is weak that allows the use of perturbation theory. At low energy, quarks and gluons are confined into   hadrons. The large coupling constant prohibits a direct application of perturbative expansions. For a high energy process with generic hard scattering, one often uses the factorization to separate the high-energy and low-energy degrees of freedoms.
The factorization approach has been widely applied to heavy meson decays~\cite{Beneke:1999br,Bauer:2000ew,Bauer:2000yr,Beneke:2003pa,Keum:2000ph,Keum:2000wi,Lu:2000em,Lu:2000hj,Kurimoto:2001zj}, in which the long-distance contributions are parametrized in terms of the low energy inputs, mostly the light-cone distribution amplitudes. For heavy baryon decays, the factorization analysis is much more involved due to the lack of knowledge on low-energy inputs and the complicated hard-scattering kernels, and see Refs.~\cite{Wang:2011uv,Feldmann:2011xf,Mannel:2011xg,Ali:2012pn,Detmold:2012vy,Bell:2013tfa} for some recent discussions.

In heavy quark decays, the flavor SU(3) symmetry is an useful tool~\cite{Savage:1989ub,Gronau:1994rj,Grinstein:1996us,He:1998rq,Deshpande:2000jp,He:2000dg,Deshpande:1994ii,Gronau:2000zy,He:2000ys,Fu:2002nr,Egolf:2002nk,Chiang:2003pm,Chiang:2004nm,Chiang:2006ih,Li:2007bh,Wang:2008rk,Chiang:2008zb,Cheng:2014rfa,He:2014xha,He:2015fwa,He:2015fsa,Hsiao:2015iiu,He:2016xvd,Chua:2016aqy,He:2017fln,Lu:2016ogy,Cheng:2016ejf,Cheng:2012xb,Cheng:2011qh,Cheng:1993kp,Cheng:1993ah,Chau:1993ec,Chau:1991gx,Li:2012cfa,Li:2013xsa}.
There are a few  advantages to adopt the SU(3) symmetry.
First once the branching fractions for a few decay channels are measured, the flavor SU(3) symmetry   offers an opportunity  to obtain the knowledge on the related channels. Secondly, the investigation of a few related decay channels can allow one to examine the CKM parameters   with the help of SU(3)  symmetry. Thirdly, when enough data is available, one may use the data to extract the SU(3) irreducible amplitudes. These amplitudes are expected to calculable in different  factorization approaches, and  can then be used to examine the factorization schemes themselves.  Thus
in this paper we will use the flavor SU(3) symmetry and analyze various decays of doubly heavy baryons.

The rest of this paper is organized as follows. In Sec.~\ref{sec:particle_multiplet}, we will collect the representations for the particle multiplets in the SU(3) symmetry.  In Sec.~\ref{sec:semileptonic}, we will analyze the semileptonic decays of the doubly-heavy baryons. The nonleptonic decays of doubly-charmed baryons, doubly-bottom baryons and the baryons with $b,c$ quarks are investigated in Sec.~\ref{sec:ccq_nonleptonic},~\ref{sec:bbq_nonleptonic} and Sec.~\ref{sec:bcq_nonleptonic}, respectively. The last section contains a brief summary.

\section{Particle Multiplets}
\label{sec:particle_multiplet}

In this section, we will collect the representations for the multiplets of the flavor SU(3) group.

\begin{table*}[!htb]
\footnotesize
\caption{Quantum numbers  for the ground state of doubly heavy baryons. The light quark $q$ corresponds to $u,d$ quark. The $J^P$ denotes the total spin and parity of the baryons.   The label $S_{h}^{\pi}$ corresponds to the spin of the heavy quark system.  }\label{tab:JPC}
\begin{center}
\begin{tabular}{cccc|cccccc} \hline \hline
Baryon      & Quark Content  &  $S_h^\pi$  &$J^P$   & Baryon & Quark Content &   $S_h^\pi$  &$J^P$   \\ \hline
$\Xi_{cc}$ & $\{cc\}q$  & $1^+$ & $1/2^+$ &   $\Xi_{bb}$ & $\{bb\}q$  & $1^+$ & $1/2^+$ & \\
$\Xi_{cc}^*$ & $\{cc\}q$  & $1^+$ & $3/2^+$ &   $\Xi_{bb}^*$ & $\{bb\}q$  & $1^+$ & $3/2^+$ & \\ \hline
$\Omega_{cc}$ & $\{cc\}s$  & $1^+$ & $1/2^+$ &   $\Omega_{bb}$ & $\{bb\}s$  & $1^+$ & $1/2^+$ & \\
$\Omega_{cc}^*$ & $\{cc\}s$  & $1^+$ & $3/2^+$ &   $\Omega_{bb}^*$ & $\{bb\}s$  & $1^+$ & $3/2^+$ &  \\ \hline
$\Xi_{bc}'$ & $\{bc\}q$  & $0^+$ & $1/2^+$ &   $\Omega_{bc}'$ & $\{bc\}s$  & $0^+$ & $1/2^+$ & \\
$\Xi_{bc}$ & $\{bc\}q$  & $1^+$ & $1/2^+$ &   $\Omega_{bc}$ & $\{bc\}s$  & $1^+$ & $1/2^+$ & \\
$\Xi_{bc}^*$ & $\{bc\}q$  & $1^+$ & $3/2^+$ &   $\Omega_{bc}^*$ & $\{bc\}s$  & $1^+$ & $3/2^+$ &
 \\ \hline \hline
\end{tabular}
\end{center}
\end{table*}

\begin{figure}
\includegraphics[width=0.5\columnwidth]{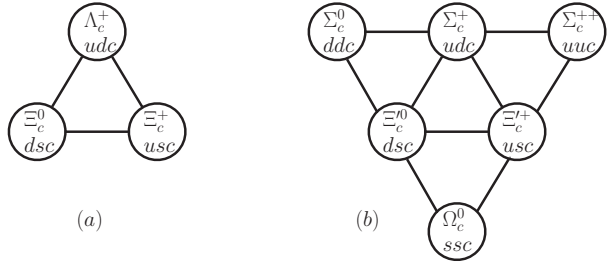}
\caption{Anti-triplets (panel a) and sextets (panel b) of charmed baryons with one charm quark and two light quarks.    }
\label{fig:one_heavy}
\end{figure}

Quantum numbers of the doubly heavy baryons are derived from the quark model~\cite{Olive:2016xmw} and  are given in Table~\ref{tab:JPC}.
These  baryons can form an SU(3) triplet which are expressed as:
\begin{eqnarray}
 T_{cc}  = \left(\begin{array}{c}  \Xi^{++}_{cc}(ccu)  \\  \Xi^+_{cc}(ccd)  \\  \Omega^+_{cc}(ccs)
\end{array}\right)\,,\;\;
  T_{bc} = \left(\begin{array}{c}  \Xi^+_{bc}(bcu)  \\  \Xi^0_{bc}(bcd)  \\  \Omega^0_{bc}(bcs)
\end{array}\right)\,,\;\;
  T_{bb} = \left(\begin{array}{c}  \Xi^0_{bb}(bbu)  \\  \Xi^-_{bb}(bbd)  \\  \Omega^-_{bb}(bbs)
\end{array}\right).
\end{eqnarray}

The singly charmed baryons  can form an anti-triplet or sextet as shown in Fig.~\ref{fig:one_heavy}. In the anti-triplet case, we have the matrix expression:
\begin{eqnarray}
 T_{\bf{c\bar 3}}= \left(\begin{array}{ccc} 0 & \Lambda_c^+  &  \Xi_c^+  \\ -\Lambda_c^+ & 0 & \Xi_c^0 \\ -\Xi_c^+   &  -\Xi_c^0  & 0
  \end{array} \right)\,.
\end{eqnarray}
For the sextet, we have the multiplet:
\begin{eqnarray}
 T_{\bf{c6}} = \left(\begin{array}{ccc} \Sigma_c^{++} &  \frac{1}{\sqrt{2}}\Sigma_c^+   & \frac{1}{\sqrt{2}} \Xi_c^{\prime+}\\
  \frac{1}{\sqrt{2}}\Sigma_c^+& \Sigma_c^{0} & \frac{1}{\sqrt{2}} \Xi_c^{\prime0} \\
  \frac{1}{\sqrt{2}} \Xi_c^{\prime+}   &  \frac{1}{\sqrt{2}} \Xi_c^{\prime0}  & \Omega_c^0
  \end{array} \right)\,.
\end{eqnarray}
This is   similar for   baryons with a bottom quark.

The light baryons form  an SU(3) octet and a decuplet. The octet has the expression:
\begin{eqnarray}
T_8= \left(\begin{array}{ccc} \frac{1}{\sqrt{2}}\Sigma^0+\frac{1}{\sqrt{6}}\Lambda^0 & \Sigma^+  &  p  \\ \Sigma^-  &  -\frac{1}{\sqrt{2}}\Sigma^0+\frac{1}{\sqrt{6}}\Lambda^0 & n \\ \Xi^-   & \Xi^0  & -\sqrt{\frac{2}{3}}\Lambda^0
  \end{array} \right),
\end{eqnarray}
while the light baryon decuplet is given as
\begin{eqnarray}
(T_{10})^{111} &=&  \Delta^{++},\;\;\; (T_{10})^{112}= (T_{10})^{121}=(T_{10})^{211}= \frac{1}{\sqrt3} \Delta^+,\nonumber\\
(T_{10})^{222} &=&  \Delta^{-},\;\;\; (T_{10})^{122}= (T_{10})^{212}=(T_{10})^{221}= \frac{1}{\sqrt3} \Delta^0, \nonumber\\
(T_{10})^{113} &=& (T_{10})^{131}=(T_{10})^{311}= \frac{1}{\sqrt3} \Sigma^{\prime+},\;\;(T_{10})^{223} = (T_{10})^{232}=(T_{10})^{322}= \frac{1}{\sqrt3} \Sigma^{\prime-},\nonumber\\
(T_{10})^{123} &=& (T_{10})^{132}=(T_{10})^{213}=(T_{10})^{231}=(T_{10})^{312}=(T_{10})^{321}= \frac{1}{\sqrt6} \Sigma^{\prime0},\nonumber\\
(T_{10})^{133} &=& (T_{10})^{313}=(T_{10})^{331}= \frac{1}{\sqrt3} \Xi^{\prime0},\;\;(T_{10})^{233} = (T_{10})^{323}=(T_{10})^{332}= \frac{1}{\sqrt3}  \Xi^{\prime-}, \nonumber\\
(T_{10})^{333}&=& \Omega^-.
\end{eqnarray}

In the meson sector,
the light pseudo-scalar meson is an octet, which can be written as:
\begin{eqnarray}
 M_{8}=\begin{pmatrix}
 \frac{\pi^0}{\sqrt{2}}+\frac{\eta}{\sqrt{6}}  &\pi^+ & K^+\\
 \pi^-&-\frac{\pi^0}{\sqrt{2}}+\frac{\eta}{\sqrt{6}}&{K^0}\\
 K^-&\bar K^0 &-2\frac{\eta}{\sqrt{6}}
 \end{pmatrix},
\end{eqnarray}
and  we will not consider the flavor singlet $\eta_1$ in this paper.
The following analysis is also applicable to the vector meson octet and other light mesons.
The  charmed meson forms an SU(3) anti-triplet:
\begin{eqnarray}
D_i=\left(\begin{array}{ccc} D^0, & D^+, & D^+_s  \end{array} \right),
\end{eqnarray}
and the anti-charmed meson forms an SU(3) triplet:
\begin{eqnarray}
\overline D^i=\left(\begin{array}{ccc}\overline D^0, & D^-, & D^-_s  \end{array} \right).
\end{eqnarray}
The above two SU(3) triplets are also applicable to the bottom mesons.

\section{Semi-Leptonic decays}
\label{sec:semileptonic}

\subsection{$\Xi_{cc}$ and $\Omega_{cc}$ decays}

\begin{figure}
\includegraphics[width=0.4\columnwidth]{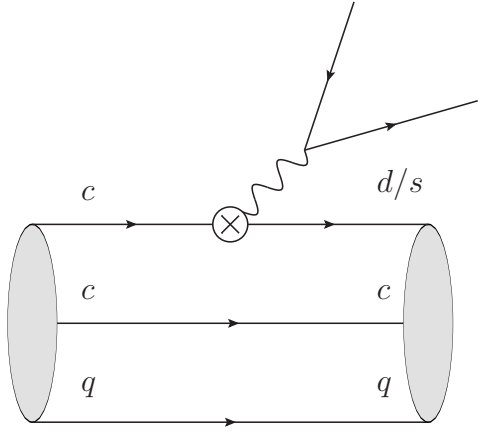}
\caption{Feynman diagrams for semileptonic decays of $\Xi_{cc}$ and $\Omega_{cc}$.  }
\label{fig:ccq_semi}
\end{figure}

The $c\to q\bar\ell\nu$ transition is induced by the effective Hamiltonian:
\begin{eqnarray}
{\cal H}_{eff}&=&\frac{G_F}{\sqrt2} \left[V_{cq}^* \bar q  \gamma^\mu(1-\gamma_5)c \bar \nu\gamma_\mu(1-\gamma_5) l\right] +h.c.,
\end{eqnarray}
where $q=d,s$ and the $V_{cd}$ and  $V_{cs}$    are   CKM matrix elements. The heavy-to-light quark operators
 will form an SU(3) triplet,  denoted as $H_{  3}$ with the components $(H_{  3})^1=0,~(H_{  3})^2=V_{cd}^*,~(H_{  3})^3=V_{cs}^*$.
At the hadron level, the effective Hamiltonian  for   decays of $\Xi_{cc}$ and $\Omega_{cc}$ into a singly charmed baryon is constructed as:
\begin{eqnarray}
  {\cal H}_{\rm{eff}}= a_1 (T_{cc})^i (H_{  3})^j  (\overline T_{\bf{c\bar 3}})_{[ij]}~\bar\nu_l l+   a_2 (T_{cc})^i (H_{  3})^j  (\overline T_{\bf{c 6}})_{\{ij\}} \bar\nu_l l\,.\label{eq:ccq_semi}
\end{eqnarray}
Here the $a_1$ and $a_2$ are   SU(3) irreducible nonperturbative   amplitudes.  Feynman diagrams for these decays are given in Fig.~\ref{fig:ccq_semi}.

The decay amplitudes for different channels can be deduced from the  Hamiltonian in Eq.~\eqref{eq:ccq_semi}, and given in Tab.~\ref{tab:ccq_semi}.  From these amplitudes, we can find the relations for   decay widths in the SU(3) symmetry limit:
\begin{eqnarray}\label{ccq_semi_relations1}
\Gamma(\Xi_{cc}^{++}\to\Lambda_c^+l^+\nu)&=& { }\Gamma(\Omega_{cc}^{+}\to\Xi_c^0l^+\nu)= \frac{|V_{cd}|^2}{|V_{cs}|^2}\Gamma(\Xi_{cc}^{++}\to\Xi_c^+l^+\nu), \\
\label{ccq_semi_relations2}
 \Gamma(\Xi_{cc}^{++}\to\Xi_c^+l^+\nu)&=  &\Gamma(\Xi_{cc}^{+}\to\Xi_c^0l^+\nu), \\
 \label{ccq_semi_relations3}
\Gamma(\Xi_{cc}^{++}\to\Sigma_{c}^{+}l^+\nu)&=& { }\Gamma(\Omega_{cc}^{+}\to\Xi_{c}^{\prime0}l^+\nu)= \frac{1}{2}\Gamma(\Xi_{cc}^{+}\to\Sigma_{c}^{0}l^+\nu)= \frac{|V_{cd}|^2}{|V_{cs}|^2}\Gamma(\Xi_{cc}^{++}\to\Xi_{c}^{\prime+}l^+\nu),\\
\label{ccq_semi_relations4}
\Gamma(\Xi_{cc}^{++}\to\Xi_{c}^{\prime+}l^+\nu)&=& { }\Gamma(\Xi_{cc}^{+}\to\Xi_{c}^{\prime0}l^+\nu)=\frac{1}{2}\Gamma(\Omega_{cc}^{+}\to\Omega_{c}^{0}l^+\nu).
\end{eqnarray}

Recently, inspired by the LHCb observation of $\Xi_{cc}$,  the weak decays of doubly heavy baryons have been studied  in Ref.~\cite{Wang:2017mqp},  where  the authors  first derived the hadronic form factors for these transitions in the light-front approach and then applied  the results  to predict the partial widths for the semi-leptonic and non-leptonic decays of doubly heavy baryons. The SU(3) symmetry can be confronted with   these results. We should note that the same comparison in semileptonic $\Xi_{bb}$, $\Omega_{bb}$ and $\Xi_{bc}$, $\Omega_{bc}$ decays and in non-leptonic decays of doubly heavy baryons can also be  made. Compared to these explicit model calculations, we found that the SU(3) symmetry works well for the  bottom quark decays, while the symmetry breaking effects are sizable for the charm quark decays, largely due to the phase-space differences.

\begin{table}
\caption{SU(3) amplitudes for doubly charmed baryons $\Xi_{cc}$ and $\Omega_{cc}$ decays into a singly charmed baryon.  }\label{tab:ccq_semi}
\begin{tabular}{|cc|cc|c|c|c|c}\hline\hline
channel & amplitude &channel & amplitude \\\hline
 $\Xi_{cc}^{++}\to \Sigma_{c}^{+}l^+\nu $ & $ \frac{a_2 V_{cd}^*}{\sqrt{2}}$ & $\Xi_{cc}^{++}\to \Lambda_c^+l^+\nu $ & $ a_1 V_{cd}^*$\\\hline
$\Xi_{cc}^{++}\to \Xi_{c}^{\prime+}l^+\nu $ & $ \frac{a_2 V_{cs}^*}{\sqrt{2}}$& $\Xi_{cc}^{++}\to \Xi_c^+l^+\nu $ & $ a_1 V_{cs}^*$   \\\hline
$\Xi_{cc}^{+}\to \Sigma_{c}^{0}l^+\nu $ & $ a_2 V_{cd}^*$& $\Xi_{cc}^{+}\to \Xi_c^0l^+\nu $ & $ a_1 V_{cs}^*$   \\\hline
$\Xi_{cc}^{+}\to \Xi_{c}^{\prime0}l^+\nu $ & $ \frac{a_2 V_{cs}^*}{\sqrt{2}}$ & $\Omega_{cc}^{+}\to \Xi_c^0l^+\nu $ & $ -a_1 V_{cd}^*$ \\\hline
$\Omega_{cc}^{+}\to \Xi_{c}^{\prime0}l^+\nu $ & $ \frac{a_2 V_{cd}^*}{\sqrt{2}}$&&\\\hline
$\Omega_{cc}^{+}\to \Omega_{c}^{0}l^+\nu $ & $ a_2 V_{cs}^*$&&\\\hline\hline
\end{tabular}
\end{table}

\subsection{Semileptonic $\Xi_{bb}$ and $\Omega_{bb}$ decays}

\begin{figure}
\includegraphics[width=0.4\columnwidth]{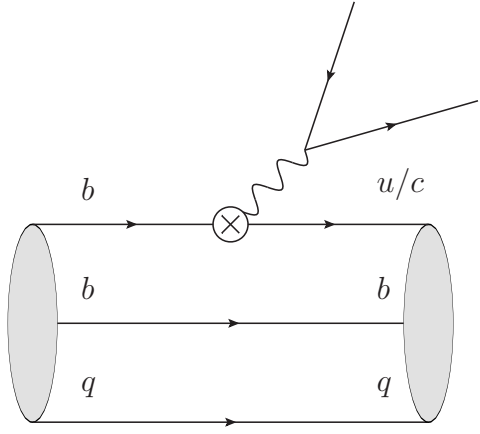}
\caption{Feynman diagrams for semileptonic decays of $\Xi_{bb}$ and $\Omega_{bb}$.  }
\label{fig:bbq_semi}
\end{figure}

The $b$ quark decay is controlled by the   Hamiltonian
\begin{eqnarray}
 {\cal H}_{eff} &=& \frac{G_F}{\sqrt2} \left[V_{q'b} \bar q' \gamma^\mu(1-\gamma_5)b \bar  l\gamma_\mu(1-\gamma_5) \nu\right] +h.c.,
\end{eqnarray}
with $q'=u,c$.  The $b\to c$ transition is an SU(3) singlet, while the $b\to u$ transition forms an SU(3) triplet $H_{3}'$ with $(H_3')^1=1$ and $(H_3')^{2,3}=0$.  Feynman diagrams can be obtained from Fig.~\ref{fig:ccq_semi} by replacing the $c$ quark by $b$ quark, and  the final $d/s$ quarks   replaced by  the $c/u$ quark.   The hadron level  Hamiltonian for  semileptonic $\Xi_{bb}$ and $\Omega_{bb}$ decays  is constructed as
\begin{eqnarray}
  H_{\rm{eff}}= a_3 (T_{bb})^i (\overline T_{bc})_i~\bar l \nu_l \,+ a_4 (T_{bb})^i (H_{  3}')^j  (\overline T_{\bf{b\bar 3}})_{[ij]}~\bar l\nu_l + a_5 (T_{bb})^i  (H_{  3}')^j  (\overline T_{\bf{b6}})_{\{ij\}}~\bar l \nu_l. \label{eq:bbq_semi}
\end{eqnarray}
 Feynman diagrams for these decays are given in Fig.~\ref{fig:bbq_semi}.
The decay amplitudes can be deduced from this Hamiltonian, and the results are given in Tab.~\ref{tab:bbq_semi}. It leads to the relations for decay widths:
\begin{eqnarray}
\Gamma(\Xi_{bb}^{0}\to\Xi_{bc}^{+}l^-\bar\nu)&=& { }\Gamma(\Xi_{bb}^{-}\to\Xi_{bc}^{0}l^-\bar\nu) = { }\Gamma(\Omega_{bb}^{-}\to\Omega_{bc}^{0}l^-\bar\nu),\\
\Gamma(\Xi_{bb}^{-}\to\Lambda_b^0l^-\bar\nu)&=& { }\Gamma(\Omega_{bb}^{-}\to\Xi_b^0l^-\bar\nu),\\
\Gamma(\Xi_{bb}^{0}\to\Sigma_{b}^{+}l^-\bar\nu)&=& 2\Gamma(\Xi_{bb}^{-}\to\Sigma_{b}^{0}l^-\bar\nu) = 2\Gamma(\Omega_{bb}^{-}\to\Xi_{b}^{\prime0}l^-\bar\nu).
\end{eqnarray}

  \begin{table}
\caption{Similar with Tab.~\ref{tab:ccq_semi} but for  the doubly bottom baryons.  }\label{tab:bbq_semi}\begin{tabular}{|c|c|c|c|c|c|c|c}\hline\hline
channel & amplitude &channel & amplitude  \\\hline
$\Xi_{bb}^{-}\to \Lambda_b^0l^-\bar\nu $ & $ -a_4 V_{\text{ub}}$ & $\Xi_{bb}^{0}\to \Xi_{bc}^{+}l^-\bar\nu $ & $ a_3 V_{\text{cb}}$\\\hline
$\Omega_{bb}^{-}\to \Xi_b^0l^-\bar\nu $ & $ -a_4 V_{\text{ub}}$& $\Xi_{bb}^{-}\to \Xi_{bc}^{0}l^-\bar\nu $ & $ a_3 V_{\text{cb}}$\\\hline
$\Xi_{bb}^{0}\to \Sigma_{b}^{+}l^-\bar\nu $ & $ a_5 V_{\text{ub}}$& $\Omega_{bb}^{-}\to \Omega_{bc}^{0}l^-\bar\nu $ & $ a_3 V_{\text{cb}}$\\\hline
$\Xi_{bb}^{-}\to \Sigma_{b}^{0}l^-\bar\nu $ & $ \frac{a_5 V_{\text{ub}}}{\sqrt{2}}$&&\\\hline
$\Omega_{bb}^{-}\to \Xi_{b}^{\prime0}l^-\bar\nu $ & $ \frac{a_5 V_{\text{ub}}}{\sqrt{2}}$&&\\\hline\hline
\end{tabular}
\end{table}


\subsection{Semileptonic $\Xi_{bc}$ and $\Omega_{bc}$ decays }

The effective Hamiltonian for semileptonic $\Xi_{bc}$ and $\Omega_{bc}$ decays is given as:
\begin{eqnarray}
  {\cal H}_{\rm{eff}} &=& a_6 (T_{bc})^i (\overline T_{cc})_i~\bar l \nu_l \,+a_7 (T_{bc})^i (H_{  3}')^j  (\overline T_{\bf{c\bar 3}})_{[ij]}~\bar\nu_l l +   a_{8} (T_{bc})^i (H_{  3}')^j  (\overline T_{\bf{c 6}})_{\{ij\}}~\bar\nu_l l \nonumber\\
  && + a_{9} (T_{bc})^i (H_{  3})^j  (\overline T_{\bf{b\bar 3}})_{[ij]}~\bar l\nu_l + a_{10} (T_{bc})^i  ( H_{  3})^j  (\overline T_{\bf{b6}})_{\{ij\}}~\bar l \nu_l.
\end{eqnarray}
In this equation, we have included both charm quark and bottom quark decays. Decay amplitudes for different channels are obtained by expanding the above Hamiltonian and  are collected in Tab.~\ref{tab:bcq_semi}.

  \begin{table}
\caption{Similar with Tab.~\ref{tab:ccq_semi} but for  the doubly-heavy $bcq$ baryons. }\label{tab:bcq_semi}\begin{tabular}{|c c|c c|c|c|c|c}\hline\hline
channel & amplitude&channel & amplitude \\\hline
$\Xi_{bc}^{+}\to \Lambda_b^0l^+\nu $ & $ a_9 V_{cd}^*$& $\Xi_{bc}^{+}\to \Xi_{cc}^{++}l^-\bar\nu $ & $ a_6 V_{{cb}}$\\\hline
$\Xi_{bc}^{+}\to \Xi_b^0l^+\nu $ & $ a_9 V_{cs}^*$& $\Xi_{bc}^{0}\to \Xi_{cc}^{+}l^-\bar\nu $ & $ a_6 V_{{cb}}$\\\hline
$\Xi_{bc}^{0}\to \Xi_b^-l^+\nu $ & $ a_9 V_{cs}^*$& $\Omega_{bc}^{0}\to \Omega_{cc}^{+}l^-\bar\nu $ & $ a_6 V_{{cb}}$\\\hline
$\Omega_{bc}^{0}\to \Xi_b^-l^+\nu $ & $ -a_9 V_{cd}^*$& $\Xi_{bc}^{0}\to \Lambda_c^+l^-\bar\nu $ & $ -a_7 V_{{ub}}$\\\hline
$\Xi_{bc}^{+}\to \Sigma_{b}^{0}l^+\nu $ & $ \frac{a_{10} V_{cd}^*}{\sqrt{2}}$& $\Omega_{bc}^{0}\to \Xi_c^+l^-\bar\nu $ & $ -a_7 V_{{ub}}$\\\hline
$\Xi_{bc}^{+}\to \Xi_{b}^{\prime0}l^+\nu $ & $ \frac{a_{10} V_{cs}^*}{\sqrt{2}}$& $\Xi_{bc}^{+}\to \Sigma_{c}^{++}l^-\bar\nu $ & $ a_8 V_{{ub}}$\\\hline
$\Xi_{bc}^{0}\to \Sigma_{b}^{-}l^+\nu $ & $ a_{10} V_{cd}^*$& $\Xi_{bc}^{0}\to \Sigma_{c}^{+}l^-\bar\nu $ & $ \frac{a_8 V_{{ub}}}{\sqrt{2}}$\\\hline
$\Xi_{bc}^{0}\to \Xi_{b}^{\prime-}l^+\nu $ & $ \frac{a_{10} V_{cs}^*}{\sqrt{2}}$&  $\Omega_{bc}^{0}\to \Xi_{c}^{\prime+}l^-\bar\nu $ & $ \frac{a_8 V_{{ub}}}{\sqrt{2}}$  \\\hline
$\Omega_{bc}^{0}\to \Xi_{b}^{\prime-}l^+\nu $ & $ \frac{a_{10} V_{cd}^*}{\sqrt{2}}$ &&\\\hline
$\Omega_{bc}^{0}\to \Omega_{b}^{-}l^+\nu $ & $ a_{10} V_{cs}^*$ &&\\\hline\hline
\end{tabular}
\end{table}

Apparently, the $\Xi_{bc}$ and $\Omega_{bc}$ decay amplitudes can be obtained by the ones for $T_{cc}$ and $T_{bb}$ decays.
For the charm quark decays, one would derive the results with the replacement, $T_{cc}\to T_{bc}$, $T_c\to T_b$. The replacement in bottom quark decays is $T_{bb}\to T_{bc}$, $T_b\to T_{c}$. Thus we have the following relations for decay widths:
\begin{eqnarray}
\Gamma(\Xi_{bc}^{+}\to\Lambda_b^0 l^+\nu)&=& { }\Gamma(\Omega_{bc}^{0}\to\Xi_b^- l^+\nu)= \frac{|V_{cd}|^2}{|V_{cs}|^2}\Gamma(\Xi_{bc}^{+}\to\Xi_b^0 l^+\nu), \\
 \Gamma(\Xi_{bc}^{+}\to\Xi_b^0 l^+\nu)&=  &\Gamma(\Xi_{bc}^{0}\to\Xi_b^- l^+\nu), \\
\Gamma(\Xi_{bc}^{+}\to\Sigma_{b}^{0}l^+\nu)&=& { }\Gamma(\Omega_{bc}^{0}\to\Xi_{b}^{\prime-}l^+\nu)= \frac{1}{2}\Gamma(\Xi_{bc}^{0}\to\Sigma_{b}^{-}l^+\nu)= \frac{|V_{cd}|^2}{|V_{cs}|^2}\Gamma(\Xi_{bc}^{+}\to\Xi_{b}^{\prime0}l^+\nu),\\
\Gamma(\Xi_{bc}^{+}\to\Xi_{b}^{\prime0}l^+\nu)&=& { }\Gamma(\Xi_{bc}^{0}\to\Xi_{b}^{\prime-}l^+\nu)=\frac{1}{2}\Gamma(\Omega_{bc}^{0}\to\Omega_{b}^{-}l^+\nu),\\
\Gamma(\Xi_{bc}^{+}\to\Xi_{cc}^{++}l^-\bar\nu)&=& { }\Gamma(\Xi_{bc}^{0}\to\Xi_{cc}^{+}l^-\bar\nu) = { }\Gamma(\Omega_{bc}^{0}\to\Omega_{cc}^{+}l^-\bar\nu),\\
\Gamma(\Xi_{bc}^{0}\to\Lambda_c^+l^-\bar\nu)&=& { }\Gamma(\Omega_{bc}^{0}\to\Xi_c^+l^-\bar\nu),\\
\Gamma(\Xi_{bc}^{+}\to\Sigma_{c}^{++}l^-\bar\nu)&=& 2\Gamma(\Xi_{bc}^{0}\to\Sigma_{c}^{+}l^-\bar\nu) = 2\Gamma(\Omega_{bc}^{0}\to\Xi_{c}^{\prime+}l^-\bar\nu).
\end{eqnarray}


\section{Non-Leptonic $\Xi_{cc}$ and $\Omega_{cc}$ decays}
\label{sec:ccq_nonleptonic}

Usually the charm quark decays into light quarks are categorized into three groups: Cabibbo allowed, singly Cabibbo suppressed, and doubly Cabibbo suppressed:
\begin{eqnarray}
 c\to s \bar d u,  \;\;\; c\to u \bar dd/\bar ss, \;\;\; c\to  d \bar s u.
\end{eqnarray}
The tree  operators    transform
under the flavor SU(3) symmetry as ${\bf  3}\otimes {\bf\bar 3}\otimes {\bf
3}={\bf  3}\oplus {\bf  3}\oplus {\bf\bar 6}\oplus {\bf {15}}$. So the Hamiltonian  can
be decomposed in terms of a vector $(H_3)$, a traceless
tensor antisymmetric in upper indices, $H_{\bf\overline6}$, and a
traceless tensor symmetric in   upper indices,
$H_{\bf {15}}$. As we will show in the following, the vector representation $H_3$ will vanishes as an approximation.

For the $c\to s  u \bar d$ transition, we have
\begin{eqnarray}
(H_{\overline 6})^{31}_2=-(H_{\overline 6})^{13}_2=1,\;\;\;
 (H_{15})^{31}_2= (H_{15})^{13}_2=1,\label{eq:H3615_c_allowed}
\end{eqnarray}
while for the $c\to d  u \bar s$ transition which is doubly Cabibbo suppressed, we have
\begin{eqnarray}
(H_{\overline 6})^{21}_3=-(H_{\overline 6})^{12}_3=\sin^2\theta_C,\;\;
 (H_{15})^{21}_3= (H_{15})^{12}_3=\sin^2\theta_C. \label{eq:H3615_c_doubly_suprressed}
\end{eqnarray}

For  the transition $c\to u \bar dd$, we have
\begin{eqnarray}
 (H_{3})^1=1,\;\;\;(H_{\overline 6})^{21}_2=-(H_{\overline 6})^{12}_2=(H_{\overline 6})^{13}_3=-(H_{\overline 6})^{31}_3=\frac{1}{2},\nonumber\\
 \frac{1}{3}(H_{15})^{21}_2= \frac{1}{3}(H_{15})^{12}_2=-\frac{1}{2}(H_{15})^{11}_1=
 -(H_{15})^{13}_3=-(H_{15})^{31}_3=\frac{1}{4},\label{eq:H3615_cc_d}
\end{eqnarray}
with all other remaining entries zero.  The overall factor is $V_{cd}^*V_{ud} \simeq -\sin(\theta_C)$. While
for  the transition $c\to u \bar ss$, we have
\begin{eqnarray}
 (H_{3})^1=1,\;\;\;(H_{\overline 6})^{31}_3=-(H_{\overline 6})^{13}_3=(H_{\overline 6})^{12}_2=-(H_{\overline 6})^{21}_2=\frac{1}{2},\nonumber\\
 \frac{1}{3}(H_{15})^{31}_3= \frac{1}{3}(H_{15})^{13}_3=-\frac{1}{2}(H_{15})^{11}_1=
 -(H_{15})^{12}_2=-(H_{15})^{21}_2=\frac{1}{4},\label{eq:H3615_cc_s}
\end{eqnarray}
with all other remaining entries zero.  The overall factor is $V_{cs}^*V_{us} \simeq \sin(\theta_C)$.
With both the $c\to u \bar dd$ and $c\to u \bar ss$, the singly Cabibbo-suppressed channel has the following effective Hamiltonian:
\begin{eqnarray}
(H_{\overline 6})^{31}_3 =-(H_{\overline 6})^{13}_3 =(H_{\overline 6})^{12}_2 =-(H_{\overline 6})^{21}_2 =\sin(\theta_C),\nonumber\\
 (H_{15})^{31}_3= (H_{15})^{13}_3=-(H_{15})^{12}_2=-(H_{15})^{21}_2= \sin(\theta_C).\label{eq:H3615_cc_singly_suppressed}
\end{eqnarray}

\subsection{Decays into a charmed baryon and a light meson}


\begin{table}
 \caption{Doubly charmed baryons decays into a $cqq$ (antitriplet) and  a light meson.}\label{tab:ccq_cqq3_qqbar}\begin{tabular}{|c  c|c  c|c|c|c|c|c|c}\hline\hline
channel & amplitude & channel & amplitude\\\hline
$\Xi_{cc}^{++}\to \Xi_c^+  \pi^+  $ & $ b_3-2 b_4+b_6$ & $\Xi_{cc}^{++}\to \Lambda_c^+  \pi^+  $ & $ \left(b_3-2 b_4+b_6\right) (-\sin(\theta_C))$\\\hline
$\Xi_{cc}^{+}\to \Lambda_c^+  \overline K^0  $ & $ b_3-b_5-b_6+b_7$ & $\Xi_{cc}^{++}\to \Xi_c^+  K^+  $ & $ \left(b_3-2 b_4+b_6\right) \sin(\theta_C)$\\\hline
$\Xi_{cc}^{+}\to \Xi_c^+  \pi^0  $ & $ \frac{2 b_4-b_5-b_7}{\sqrt{2}}$ & $\Xi_{cc}^{+}\to \Lambda_c^+  \pi^0  $ & $ \frac{\left(b_3-2 b_4-b_6+2 b_7\right) \sin(\theta_C)}{\sqrt{2}}$\\\hline
$\Xi_{cc}^{+}\to \Xi_c^+  \eta  $ & $ \frac{-2 b_4+b_5-3 b_7}{\sqrt{6}}$ & $\Xi_{cc}^{+}\to \Lambda_c^+  \eta  $ & $ \frac{\left(-3 b_3+2 b_4+2 b_5+3 b_6\right) \sin(\theta_C)}{\sqrt{6}}$\\\hline
$\Xi_{cc}^{+}\to \Xi_c^0  \pi^+  $ & $ b_3-b_5+b_6-b_7$ & $\Xi_{cc}^{+}\to \Xi_c^+  K^0  $ & $ \left(2 b_4-b_5+b_7\right) (-\sin(\theta_C))$\\\hline
$\Omega_{cc}^{+}\to \Xi_c^+  \overline K^0  $ & $ b_3-2 b_4-b_6$ & $\Xi_{cc}^{+}\to \Xi_c^0  K^+  $ & $ \left(b_3-b_5+b_6-b_7\right) \sin(\theta_C)$\\\hline
\Xcline{1-2}{1.2pt}
\hline$\Xi_{cc}^{++}\to \Lambda_c^+  K^+  $ & $ \left(b_3-2 b_4+b_6\right) \sin^2(\theta_C)$ & $\Omega_{cc}^{+}\to \Lambda_c^+  \overline K^0  $ & $ \left(2 b_4-b_5+b_7\right) \sin(\theta_C)$\\\hline
$\Xi_{cc}^{+}\to \Lambda_c^+  K^0  $ & $ \left(b_3-2 b_4-b_6\right) \sin^2(\theta_C)$ & $\Omega_{cc}^{+}\to \Xi_c^+  \pi^0  $ & $ \frac{\left(b_3-b_5-b_6-b_7\right) \sin(\theta_C)}{\sqrt{2}}$\\\hline
$\Omega_{cc}^{+}\to \Lambda_c^+  \pi^0  $ & $ -\sqrt{2} b_7 \sin^2(\theta_C)$ & $\Omega_{cc}^{+}\to \Xi_c^+  \eta  $ & $ \frac{\left(-3 b_3+4 b_4+b_5+3 b_6-3 b_7\right) \sin(\theta_C)}{\sqrt{6}}$\\\hline
$\Omega_{cc}^{+}\to \Lambda_c^+  \eta  $ & $ \sqrt{\frac{2}{3}} \left(2 b_4-b_5\right) \sin^2(\theta_C)$ & $\Omega_{cc}^{+}\to \Xi_c^0  \pi^+  $ & $ \left(b_3-b_5+b_6-b_7\right) \sin(\theta_C)$\\\hline
$\Omega_{cc}^{+}\to \Xi_c^+  K^0  $ & $ \left(b_3-b_5-b_6+b_7\right) \sin^2(\theta_C)$  &  &\\\hline
$\Omega_{cc}^{+}\to \Xi_c^0  K^+  $ & $ \left(b_3-b_5+b_6-b_7\right) \left(-\sin^2(\theta_C)\right)$  &  &\\\hline
\hline
\end{tabular}
\end{table}

  \begin{table}
\caption{Doubly charmed baryons decays into a $cqq$ (sextet) and  a light meson.}\label{tab:ccq_cqq6_qqbar}\begin{tabular}{|c  c|c  c|c|c|c|c|c|c}\hline\hline
channel & amplitude & channel & amplitude\\\hline
$\Xi_{cc}^{++}\to \Sigma_{c}^{++}  \overline K^0  $ & $ b_{10}-b_{13}$ & $\Xi_{cc}^{++}\to \Sigma_{c}^{++}  \pi^0  $ & $ \frac{\left(b_{10}-b_{13}\right) \sin(\theta_C)}{\sqrt{2}}$\\\hline
$\Xi_{cc}^{++}\to \Xi_{c}^{\prime+}  \pi^+  $ & $ \frac{b_{10}+2 b_{11}+b_{13}}{\sqrt{2}}$ & $\Xi_{cc}^{++}\to \Sigma_{c}^{++}  \eta  $ & $ -\sqrt{\frac{3}{2}} \left(b_{10}-b_{13}\right) \sin(\theta_C)$\\\hline
$\Xi_{cc}^{+}\to \Sigma_{c}^{++}  K^-  $ & $ b_{12}-b_{14}$ & $\Xi_{cc}^{++}\to \Sigma_{c}^{+}  \pi^+  $ & $ -\frac{\left(b_{10}+2 b_{11}+b_{13}\right) \sin(\theta_C)}{\sqrt{2}}$\\\hline
$\Xi_{cc}^{+}\to \Sigma_{c}^{+}  \overline K^0  $ & $ \frac{b_{10}+b_{12}-b_{13}-b_{14}}{\sqrt{2}}$ & $\Xi_{cc}^{++}\to \Xi_{c}^{\prime+}  K^+  $ & $ \frac{\left(b_{10}+2 b_{11}+b_{13}\right) \sin(\theta_C)}{\sqrt{2}}$\\\hline
$\Xi_{cc}^{+}\to \Xi_{c}^{\prime+}  \pi^0  $ & $ \frac{1}{2} \left(-2 b_{11}+b_{12}+b_{14}\right)$ & $\Xi_{cc}^{+}\to \Sigma_{c}^{++}  \pi^-  $ & $ \left(b_{14}-b_{12}\right) \sin(\theta_C)$\\\hline
$\Xi_{cc}^{+}\to \Xi_{c}^{\prime+}  \eta  $ & $ \frac{2 b_{11}-b_{12}+3 b_{14}}{2 \sqrt{3}}$ & $\Xi_{cc}^{+}\to \Sigma_{c}^{+}  \pi^0  $ & $ \frac{1}{2} \left(b_{10}+2 b_{11}-b_{13}-2 b_{14}\right) \sin(\theta_C)$\\\hline
$\Xi_{cc}^{+}\to \Xi_{c}^{\prime0}  \pi^+  $ & $ \frac{b_{10}+b_{12}+b_{13}+b_{14}}{\sqrt{2}}$ & $\Xi_{cc}^{+}\to \Sigma_{c}^{+}  \eta  $ & $ -\frac{\left(3 b_{10}+2 b_{11}+2 b_{12}-3 b_{13}\right) \sin(\theta_C)}{2 \sqrt{3}}$\\\hline
$\Xi_{cc}^{+}\to \Omega_{c}^{0}  K^+  $ & $ b_{12}+b_{14}$ & $\Xi_{cc}^{+}\to \Sigma_{c}^{0}  \pi^+  $ & $ \left(b_{10}+b_{12}+b_{13}+b_{14}\right) (-\sin(\theta_C))$\\\hline
$\Omega_{cc}^{+}\to \Xi_{c}^{\prime+}  \overline K^0  $ & $ \frac{b_{10}+2 b_{11}-b_{13}}{\sqrt{2}}$ & $\Xi_{cc}^{+}\to \Xi_{c}^{\prime+}  K^0  $ & $ \frac{\left(2 b_{11}-b_{12}+b_{14}\right) \sin(\theta_C)}{\sqrt{2}}$\\\hline
$\Omega_{cc}^{+}\to \Omega_{c}^{0}  \pi^+  $ & $ b_{10}+b_{13}$ & $\Xi_{cc}^{+}\to \Xi_{c}^{\prime0}  K^+  $ & $ \frac{\left(b_{10}-b_{12}+b_{13}-b_{14}\right) \sin(\theta_C)}{\sqrt{2}}$\\\hline
\Xcline{1-2}{1.2pt}
\hline$\Xi_{cc}^{++}\to \Sigma_{c}^{++}  K^0  $ & $ \left(b_{10}-b_{13}\right) \sin^2(\theta_C)$ & $\Omega_{cc}^{+}\to \Sigma_{c}^{++}  K^-  $ & $ \left(b_{12}-b_{14}\right) \sin(\theta_C)$\\\hline
$\Xi_{cc}^{++}\to \Sigma_{c}^{+}  K^+  $ & $ \frac{\left(b_{10}+2 b_{11}+b_{13}\right) \sin^2(\theta_C)}{\sqrt{2}}$ & $\Omega_{cc}^{+}\to \Sigma_{c}^{+}  \overline K^0  $ & $ -\frac{\left(2 b_{11}-b_{12}+b_{14}\right) \sin(\theta_C)}{\sqrt{2}}$\\\hline
$\Xi_{cc}^{+}\to \Sigma_{c}^{+}  K^0  $ & $ \frac{\left(b_{10}+2 b_{11}-b_{13}\right) \sin^2(\theta_C)}{\sqrt{2}}$ & $\Omega_{cc}^{+}\to \Xi_{c}^{\prime+}  \pi^0  $ & $ \frac{1}{2} \left(b_{10}+b_{12}-b_{13}+b_{14}\right) \sin(\theta_C)$\\\hline
$\Xi_{cc}^{+}\to \Sigma_{c}^{0}  K^+  $ & $ \left(b_{10}+b_{13}\right) \sin^2(\theta_C)$ & $\Omega_{cc}^{+}\to \Xi_{c}^{\prime+}  \eta  $ & $ -\frac{\left(3 b_{10}+4 b_{11}+b_{12}-3 b_{13}-3 b_{14}\right) \sin(\theta_C)}{2 \sqrt{3}}$\\\hline
$\Omega_{cc}^{+}\to \Sigma_{c}^{++}  \pi^-  $ & $ \left(b_{12}-b_{14}\right) \sin^2(\theta_C)$ & $\Omega_{cc}^{+}\to \Xi_{c}^{\prime0}  \pi^+  $ & $ -\frac{\left(b_{10}-b_{12}+b_{13}-b_{14}\right) \sin(\theta_C)}{\sqrt{2}}$\\\hline
$\Omega_{cc}^{+}\to \Sigma_{c}^{+}  \pi^0  $ & $ b_{14} \sin^2(\theta_C)$ & $\Omega_{cc}^{+}\to \Omega_{c}^{0}  K^+  $ & $ \left(b_{10}+b_{12}+b_{13}+b_{14}\right) \sin(\theta_C)$\\\hline
$\Omega_{cc}^{+}\to \Sigma_{c}^{+}  \eta  $ & $ \frac{\left(b_{12}-2 b_{11}\right) \sin^2(\theta_C)}{\sqrt{3}}$ &&\\\hline
$\Omega_{cc}^{+}\to \Sigma_{c}^{0}  \pi^+  $ & $ \left(b_{12}+b_{14}\right) \sin^2(\theta_C)$&&\\\hline
$\Omega_{cc}^{+}\to \Xi_{c}^{\prime+}  K^0  $ & $ \frac{\left(b_{10}+b_{12}-b_{13}-b_{14}\right) \sin^2(\theta_C)}{\sqrt{2}}$&&\\\hline
$\Omega_{cc}^{+}\to \Xi_{c}^{\prime0}  K^+  $ & $ \frac{\left(b_{10}+b_{12}+b_{13}+b_{14}\right) \sin^2(\theta_C)}{\sqrt{2}}$&&\\\hline
\hline
\end{tabular}
\end{table}

\begin{figure}
\includegraphics[width=0.9\columnwidth]{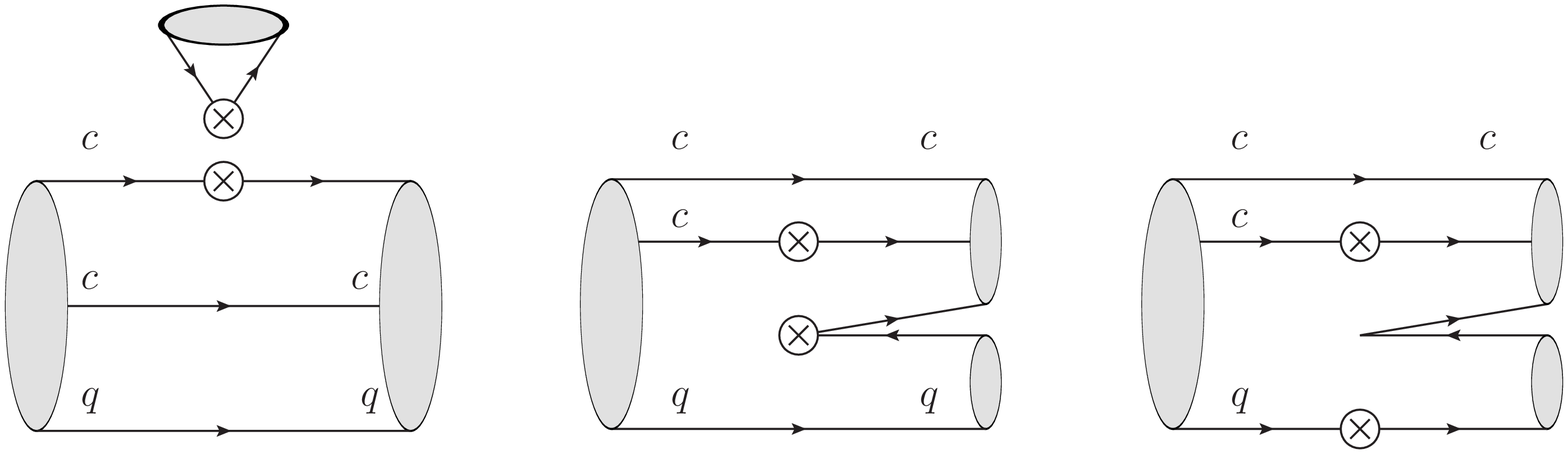}
\caption{Feynman diagrams for $\Xi_{cc}$ and $\Omega_{cc}$ decays into a charmed baryon and a light meson.  }
\label{ccq_cqqandmeson}
\end{figure}

With the above expressions, one may derive the   effective Hamiltonian for decays involving the anti-triplet heavy baryons  as
\begin{eqnarray}
 {\cal H}_{eff}&=&  b_3(T_{cc})^i  (\overline T_{c\bar 3})_{[ij]} M^{k}_{l}  (H_{\overline6})^{jl}_{k}+ b_4(T_{cc})^i  (\overline T_{c\bar 3})_{[jl]} M^{k}_{i}  (H_{\overline6})^{jl}_{k}+ b_5(T_{cc})^i  (\overline T_{c\bar 3})_{[jk]} M^{k}_{l}  (H_{\overline6})^{jl}_{i} \nonumber\\
 && +  b_6(T_{cc})^i  (\overline T_{c\bar 3})_{[ij]} M^{k}_{l}  (H_{15})^{jl}_{k}+ b_7(T_{cc})^i  (\overline T_{c\bar 3})_{[jk]} M^{k}_{l}  (H_{15})^{jl}_{i}.
\end{eqnarray}
For the sextet baryon, we have  the Hamiltonian
\begin{eqnarray}
 {\cal H}_{eff}&=&  b_{10}(T_{cc})^i  (\overline T_{c6})_{\{ij\}} M^{k}_{l}  (H_{15})^{jl}_{k}+ b_{11}(T_{cc})^i  (\overline T_{c6})_{\{jl\}} M^{k}_{i}  (H_{15})^{jl}_{k}+ b_{12}(T_{cc})^i  (\overline T_{c6})_{\{jk\}} M^{k}_{l}  (H_{15})^{jl}_{i} \nonumber\\
 && +  b_{13}(T_{cc})^i  (\overline T_{c6})_{\{ij\}} M^{k}_{l}  (H_{\overline6})^{jk}_{l}+ b_{14}(T_{cc})^i  (\overline T_{c6})_{\{jk\}} M^{k}_{l}  (H_{\overline6})^{jl}_{i}.
\end{eqnarray}
Feynman diagrams for these decays are given in Fig.~\ref{ccq_cqqandmeson}.

Expanding the above equations, we will obtain the decay amplitudes given in Tab.~\ref{tab:ccq_cqq3_qqbar} for the antitriplet baryon and Tab.~\ref{tab:ccq_cqq6_qqbar} for the sextet. Thus we have the following relations for decay widths:
\begin{eqnarray}
    \Gamma(\Xi_{cc}^{++}\to\Lambda_c^+\pi^+ )= { }\Gamma(\Xi_{cc}^{++}\to\Xi_c^+K^+ ),\\ \Gamma(\Xi_{cc}^{+}\to\Xi_c^+K^0 )= { }\Gamma(\Omega_{cc}^{+}\to\Lambda_c^+\overline K^0 ),\\ \Gamma(\Omega_{cc}^{+}\to\Xi_c^0\pi^+ )= { }\Gamma(\Xi_{cc}^{+}\to\Xi_c^0K^+ ).
     \end{eqnarray}
For the decays into the sextet, we have:
\begin{eqnarray}
    \Gamma(\Xi_{cc}^{++}\to\Sigma_{c}^{++}\pi^0 )= \frac{1}{3}\Gamma(\Xi_{cc}^{++}\to\Sigma_{c}^{++}\eta ),\\ \Gamma(\Xi_{cc}^{++}\to\Sigma_{c}^{+}\pi^+ )= { }\Gamma(\Xi_{cc}^{++}\to\Xi_{c}^{\prime+}K^+ ),\\ \Gamma(\Xi_{cc}^{+}\to\Sigma_{c}^{++}\pi^- )= { }\Gamma(\Omega_{cc}^{+}\to\Sigma_{c}^{++}K^- ),\\ \Gamma(\Xi_{cc}^{+}\to\Sigma_{c}^{0}\pi^+ )= { }\Gamma(\Omega_{cc}^{+}\to\Omega_{c}^{0}K^+ ),\\ \Gamma(\Xi_{cc}^{+}\to\Xi_{c}^{\prime+}K^0 )= { }\Gamma(\Omega_{cc}^{+}\to\Sigma_{c}^{+}\overline K^0 ),\\ \Gamma(\Omega_{cc}^{+}\to\Xi_{c}^{\prime0}\pi^+ )= { }\Gamma(\Xi_{cc}^{+}\to\Xi_{c}^{\prime0}K^+ ). \end{eqnarray}

\subsection{Decays into a light octet baryon and a charmed meson}

\begin{figure}
\includegraphics[width=0.6\columnwidth]{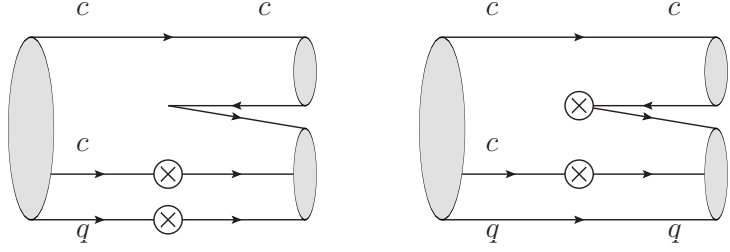}
\caption{Feynman diagrams for $\Xi_{cc}$ and $\Omega_{cc}$ decays into a light baryon and a charmed meson.  }
\label{ccq_qqq8andcharmedmeson}
\end{figure}

The effective Hamiltonian for the decays of $T_{cc}$ into a light octet baryon and a charmed meson  is given as
\begin{eqnarray}
 {\cal H}_{eff}&=&  c_4(T_{cc})^l  \overline D ^m  \epsilon_{ijk} (T_8)^{k}_{l}  (H_{6})^{ij}_m +c_5(T_{cc})^l  \overline D^m  \epsilon_{ijk} (T_8)^{k}_{m}  (H_{6})^{ij}_l +c_6(T_{cc})^l  \overline D ^i\epsilon_{ijk} (T_8)^{k}_{m}  (H_{6})^{jm}_l\nonumber\\
&&+c_7(T_{cc})^i  \overline D^l\epsilon_{ijk} (T_8)^{k}_{m}  (H_{6})^{jm}_l +c_8(T_{cc})^l  \overline D^i\epsilon_{ijk} (T_8)^{k}_{m}  (H_{15})^{jm}_l+c_9(T_{cc})^i  \overline D^l\epsilon_{ijk} (T_8)^{k}_{m}  (H_{15})^{jm}_l. \nonumber\\
\end{eqnarray}
In the above Hamiltonian we find the following relatios:
\begin{eqnarray}
(T_{cc})^l  \overline D ^m  \epsilon_{ijk} (T_8)^{k}_{l}  (H_{6})^{ij}_m= -2 (T_{cc})^i  \overline D^l\epsilon_{ijk} (T_8)^{k}_{m}  (H_{6})^{jm}_l, \nonumber\\
(T_{cc})^l  \overline D^m  \epsilon_{ijk} (T_8)^{k}_{m}  (H_{6})^{ij}_l= -2 (T_{cc})^l  \overline D ^i\epsilon_{ijk} (T_8)^{k}_{m}  (H_{6})^{jm}_l.
\end{eqnarray}
Thus two of the reduced matrix elements are not independent. In the following, we will eliminate the $c_4$ and $c_5$ and use the effective Hamiltonian:
\begin{eqnarray}
 {\cal H}_{eff}&=& c_6(T_{cc})^l  \overline D ^i\epsilon_{ijk} (T_8)^{k}_{m}  (H_{6})^{jm}_l+c_7(T_{cc})^i  \overline D^l\epsilon_{ijk} (T_8)^{k}_{m}  (H_{6})^{jm}_l\nonumber\\
&& +c_8(T_{cc})^l  \overline D^i\epsilon_{ijk} (T_8)^{k}_{m}  (H_{15})^{jm}_l+c_9(T_{cc})^i  \overline D^l\epsilon_{ijk} (T_8)^{k}_{m}  (H_{15})^{jm}_l.
\end{eqnarray}
Feynman diagrams for these decays are given in Fig.~\ref{ccq_qqq8andcharmedmeson}. Expanding the above equations, we will obtain the decay amplitudes given in Tab.~\ref{tab:ccq_qqq8_cqbar},
which leads to the relations for decay widths:
\begin{eqnarray}
    \Gamma(\Xi_{cc}^{++}\to{p}D^+)&=& { }\Gamma(\Xi_{cc}^{++}\to\Sigma^+D^+_s), \nonumber\\
    \Gamma(\Xi_{cc}^{+}\to{n}D^+)&=& { }\Gamma(\Omega_{cc}^{+}\to\Xi^0D^+_s),\nonumber\\
    \Gamma(\Omega_{cc}^{+}\to\Sigma^+D^0)&=& { }\Gamma(\Xi_{cc}^{+}\to{p}D^0).
     \end{eqnarray}

 \begin{table}
\caption{Doubly charmed baryons decays into a light baryon in the octet  and  a charmed meson.}\label{tab:ccq_qqq8_cqbar}\begin{tabular}{|c  c|c  c|c|c|c|c|c|c}\hline\hline
channel & amplitude & channel & amplitude \\\hline
$\Xi_{cc}^{++}\to \Sigma^+  D^+ $ & $-c_7-c_9$ & $\Xi_{cc}^{++}\to \Sigma^+  D^+_s $ & $ \left(-c_7-c_9\right) \sin(\theta_C)$\\\hline
$\Xi_{cc}^{+}\to \Lambda^0  D^+ $ & $ \frac{-c_6-c_7+3 c_8+3 c_9}{\sqrt{6}}$ & $\Xi_{cc}^{++}\to {p}  D^+ $ & $ \left(-c_7-c_9\right) \sin(\theta_C)$\\\hline
$\Xi_{cc}^{+}\to \Sigma^+  D^0 $ & $ -c_6-c_8$ & $\Xi_{cc}^{+}\to \Lambda^0  D^+_s $ & $ \frac{\left(2 c_6-c_7+3 c_9\right) \sin(\theta_C)}{\sqrt{6}}$\\\hline
$\Xi_{cc}^{+}\to \Sigma^0  D^+ $ & $ \frac{c_6+c_7+c_8+c_9}{\sqrt{2}}$ & $\Xi_{cc}^{+}\to \Sigma^0  D^+_s $ & $ \frac{\left(c_7+2 c_8+c_9\right) \sin(\theta_C)}{\sqrt{2}}$\\\hline
$\Xi_{cc}^{+}\to \Xi^0  D^+_s $ & $-c_6+c_8$ & $\Xi_{cc}^{+}\to {p}  D^0 $ & $ \left(-c_6-c_8\right) \sin(\theta_C)$\\\hline
$\Omega_{cc}^{+}\to \Xi^0  D^+ $ & $-c_7+c_9$ & $\Xi_{cc}^{+}\to {n}  D^+ $ & $ \left(-c_6-c_7+c_8+c_9\right) \sin(\theta_C)$\\\hline
\Xcline{1-2}{1.2pt}
\hline$\Xi_{cc}^{++}\to {p}  D^+_s $ & $ \left(c_7+c_9\right) \sin^2(\theta_C)$ & $\Omega_{cc}^{+}\to \Lambda^0  D^+ $ & $ -\frac{\left(c_6-2 c_7-3 c_8\right) \sin(\theta_C)}{\sqrt{6}}$\\\hline
$\Xi_{cc}^{+}\to {n}  D^+_s $ & $ \left(c_7-c_9\right) \sin^2(\theta_C)$ & $\Omega_{cc}^{+}\to \Sigma^+  D^0 $ & $ \left(-c_6-c_8\right) \sin(\theta_C)$\\\hline
$\Omega_{cc}^{+}\to \Lambda^0  D^+_s $ & $ \sqrt{\frac{2}{3}} \left(-c_6-c_7\right) \sin^2(\theta_C)$ &　$\Omega_{cc}^{+}\to \Sigma^0  D^+ $ & $ \frac{\left(c_6+c_8+2 c_9\right) \sin(\theta_C)}{\sqrt{2}}$\\\hline
$\Omega_{cc}^{+}\to \Sigma^0  D^+_s $ & $ -\sqrt{2} \left(c_8+c_9\right) \sin^2(\theta_C)$　& $\Omega_{cc}^{+}\to \Xi^0  D^+_s $ & $ \left(-c_6-c_7+c_8+c_9\right) \sin(\theta_C)$\\\hline
$\Omega_{cc}^{+}\to {p}  D^0 $ & $ \left(c_6+c_8\right) \sin^2(\theta_C)$ &&\\\hline
$\Omega_{cc}^{+}\to {n}  D^+ $ & $ \left(c_6-c_8\right) \sin^2(\theta_C)$ &&\\\hline
\hline
\end{tabular}
\end{table}


\subsection{Decays into a light decuplet baryon and a charmed meson }

 The effective Hamiltonian for a light decuplet in the final state is given as
\begin{eqnarray}
 {\cal H}_{eff}&=&  d_4(T_{cc})^l  \overline D ^m   (T_{10})_{ijl}  (H_{15})^{ij}_m +d_5(T_{cc})^l  \overline D^m  (T_{10})_{ijm}  (H_{15})^{ij}_l .
\end{eqnarray}
Feynman diagrams for these decays are same as Fig.~\ref{ccq_qqq8andcharmedmeson}. The corresponding decay amplitudes are given in Tab.~\ref{tab:ccq_qqq10_cqbar} and
it leads to the relations for decay widths:
\begin{eqnarray}
    \Gamma(\Xi_{cc}^{++}\to\Delta^{+}D^+)= { }\Gamma(\Xi_{cc}^{++}\to\Sigma^{\prime+}D^+_s),\\ \Gamma(\Xi_{cc}^{+}\to\Delta^{0}D^+)= { }\Gamma(\Omega_{cc}^{+}\to\Xi^{\prime0}D^+_s),\\ \Gamma(\Xi_{cc}^{+}\to\Sigma^{\prime+}D^0)= { }\Gamma(\Xi_{cc}^{+}\to\Xi^{\prime0}D^+_s),\\ \Gamma(\Omega_{cc}^{+}\to\Delta^{+}D^0)= { }\Gamma(\Omega_{cc}^{+}\to\Delta^{0}D^+),\\ \Gamma(\Omega_{cc}^{+}\to\Sigma^{\prime0}D^+)= { }\Gamma(\Xi_{cc}^{+}\to\Sigma^{\prime0}D^+_s), \\
    \Gamma(\Xi_{cc}^{++}\to\Delta^{+}D^+_s)= { }\Gamma(\Xi_{cc}^{+}\to\Delta^{0}D^+_s),\\
    \Gamma(\Xi_{cc}^{++}\to\Sigma^{\prime+}D^+)= { }\Gamma(\Omega_{cc}^{+}\to\Xi^{\prime0}D^+),\\
    \Gamma(\Xi_{cc}^{+}\to\Delta^{+}D^0)= { }\Gamma(\Omega_{cc}^{+}\to\Sigma^{\prime+}D^0).
    \end{eqnarray}

  \begin{table}
\caption{Doubly charmed baryons decays into a light baryon in the decuplet  and  a charmed meson.}\label{tab:ccq_qqq10_cqbar}\begin{tabular}{|c  c|c  c|c|c|c|c|c|c}\hline\hline
channel & amplitude & channel & amplitude\\\hline
$\Xi_{cc}^{++}\to \Sigma^{\prime+}  D^+ $ & $ \frac{2 d_4}{\sqrt{3}}$ & $\Xi_{cc}^{++}\to \Delta^{+}  D^+ $ & $ -\frac{2 d_4 \sin(\theta_C)}{\sqrt{3}}$\\\hline
$\Xi_{cc}^{+}\to \Sigma^{\prime+}  D^0 $ & $ \frac{2 d_5}{\sqrt{3}}$ & $\Xi_{cc}^{++}\to \Sigma^{\prime+}  D^+_s $ & $ \frac{2 d_4 \sin(\theta_C)}{\sqrt{3}}$\\\hline
$\Xi_{cc}^{+}\to \Sigma^{\prime0}  D^+ $ & $ \sqrt{\frac{2}{3}} \left(d_4+d_5\right)$ & $\Xi_{cc}^{+}\to \Delta^{+}  D^0 $ & $ -\frac{2 d_5 \sin(\theta_C)}{\sqrt{3}}$\\\hline
$\Xi_{cc}^{+}\to \Xi^{\prime0}  D^+_s $ & $ \frac{2 d_5}{\sqrt{3}}$ & $\Xi_{cc}^{+}\to \Delta^{0}  D^+ $ & $ -\frac{2 \left(d_4+d_5\right) \sin(\theta_C)}{\sqrt{3}}$\\\hline
$\Omega_{cc}^{+}\to \Xi^{\prime0}  D^+ $ & $ \frac{2 d_4}{\sqrt{3}}$ & $\Xi_{cc}^{+}\to \Sigma^{\prime0}  D^+_s $ & $ \sqrt{\frac{2}{3}} \left(d_4-d_5\right) \sin(\theta_C)$\\\hline
\Xcline{1-2}{1.2pt}
$\Xi_{cc}^{++}\to \Delta^{+}  D^+_s $ & $ \frac{2 d_4 \sin^2(\theta_C)}{\sqrt{3}}$ & $\Omega_{cc}^{+}\to \Sigma^{\prime+}  D^0 $ & $ \frac{2 d_5 \sin(\theta_C)}{\sqrt{3}}$\\\hline
$\Xi_{cc}^{+}\to \Delta^{0}  D^+_s $ & $ \frac{2 d_4 \sin^2(\theta_C)}{\sqrt{3}}$ & $\Omega_{cc}^{+}\to \Sigma^{\prime0}  D^+ $ & $ \sqrt{\frac{2}{3}} \left(d_5-d_4\right) \sin(\theta_C)$\\\hline
$\Omega_{cc}^{+}\to \Delta^{+}  D^0 $ & $ \frac{2 d_5 \sin^2(\theta_C)}{\sqrt{3}}$ & $\Omega_{cc}^{+}\to \Xi^{\prime0}  D^+_s $ & $ \frac{2 \left(d_4+d_5\right) \sin(\theta_C)}{\sqrt{3}}$\\\hline
$\Omega_{cc}^{+}\to \Delta^{0}  D^+ $ & $ \frac{2 d_5 \sin^2(\theta_C)}{\sqrt{3}}$ &&\\\hline
$\Omega_{cc}^{+}\to \Sigma^{\prime0}  D^+_s $ & $ \sqrt{\frac{2}{3}} \left(d_4+d_5\right) \sin^2(\theta_C)$ && \\\hline
\hline
\end{tabular}
\end{table}


In addition from the decay amplitudes, one can see that  there are relations between the widths between
Cabibbo-allowed, singly Cabibbo suppressed and doubly Cabibbo suppressed decay modes:
\begin{eqnarray}
r ^2 = \frac{\Gamma({\rm channel} \;\; 1)}{\Gamma({\rm channel}\;\; 2)}.
\end{eqnarray}
These  relations are given in Tab.~\ref{tab:relations_cabibbo1}, \ref{tab:relations_cabibbo2} and Tab.~\ref{tab:relations_cabibbo3}, respectively.

  \begin{table}[http]
\caption{ Relations for Cabibbo-allowed, singly Cabibbo suppressed and doubly Cabibbo suppressed processes in Tab.~\ref{tab:ccq_cqq3_qqbar} and Tab.~\ref{tab:ccq_qqq8_cqbar}. }\label{tab:relations_cabibbo1}
\begin{tabular}{|c|c|c||c|c|c|c|c}\hline\hline
channel 1 & channel 2 & $r$ & channel 1 & channel 2 & $r$\\\hline
$\Xi^{++}_{cc} \to  \Lambda^+_c \pi^+ $&$\Xi^{++}_{cc} \to  \Lambda^+_c K^+ $&$-\csc \left(\theta _c\right)$ &     $\Xi^+_{cc} \to  \Xi^+_c K^0 $&$\Omega^+_{cc} \to  \Lambda^+_c \bar{K}^0$&$-1$\\\hline
 $\Xi^{++}_{cc} \to  \Lambda^+_c \pi^+ $&$\Xi^{++}_{cc} \to  \Xi^+_c \pi^+ $&$-\sin \left(\theta _c\right)$ &      $\Xi^+_{cc} \to  \Xi^0_c \pi^+ $&$\Xi^+_{cc} \to  \Xi^0_c K^+ $&$\csc \left(\theta _c\right)$\\\hline
 $\Xi^{++}_{cc} \to  \Lambda^+_c \pi^+ $&$\Xi^{++}_{cc} \to  \Xi^+_c K^+ $&$-1$ &       $\Xi^+_{cc} \to  \Xi^0_c \pi^+ $&$\Omega^+_{cc} \to  \Xi^0_c \pi^+ $&$\csc \left(\theta _c\right)$\\\hline
 $\Xi^{++}_{cc} \to  \Lambda^+_c K^+ $&$\Xi^{++}_{cc} \to  \Xi^+_c \pi^+ $&$\sin ^2\left(\theta _c\right)$ &       $\Xi^+_{cc} \to  \Xi^0_c \pi^+ $&$\Omega^+_{cc} \to  \Xi^0_c K^+ $&$-\csc ^2\left(\theta _c\right)$\\\hline
  $\Xi^{++}_{cc} \to  \Lambda^+_c K^+ $&$\Xi^{++}_{cc} \to  \Xi^+_c K^+ $&$\sin \left(\theta _c\right)$ &       $\Xi^+_{cc} \to  \Xi^0_c K^+ $&$\Omega^+_{cc} \to  \Xi^0_c \pi^+ $&$1$\\\hline
  $\Xi^{++}_{cc} \to  \Xi^+_c \pi^+ $&$\Xi^{++}_{cc} \to  \Xi^+_c K^+ $&$\csc \left(\theta _c\right)$ &        $\Xi^+_{cc} \to  \Xi^0_c K^+ $&$\Omega^+_{cc} \to  \Xi^0_c K^+ $&$-\csc \left(\theta _c\right)$\\\hline
  $\Xi^+_{cc} \to  \Lambda^+_c K^0 $&$\Omega^+_{cc} \to  \Xi^+_c \bar{K}^0$&$\sin ^2\left(\theta _c\right)$ &        $\Omega^+_{cc} \to  \Xi^0_c \pi^+ $&$\Omega^+_{cc} \to  \Xi^0_c K^+ $&$-\csc \left(\theta _c\right)$\\\hline
   $\Xi^+_{cc} \to  \Lambda^+_c \bar{K}^0$&$\Omega^+_{cc} \to  \Xi^+_c K^0 $&$\csc ^2\left(\theta _c\right)$ &&&\\\hline\hline
   $\Xi^{++}_{cc} \to  p D^+_s $&$\Xi^{++}_{cc} \to  \Sigma^+ D^+ $&$-\sin ^2\left(\theta _c\right)$ &        $\Xi^+_{cc} \to  p D^0 $&$\Xi^+_{cc} \to  \Sigma^+ D^0 $&$\sin \left(\theta _c\right)$\\\hline
    $\Xi^+_{cc} \to  n D^+_s $&$\Omega^+_{cc} \to  \Xi^0D^+ $&$-\sin ^2\left(\theta _c\right)$ &        $\Omega^+_{cc} \to  \Sigma^+ D^0 $&$\Xi^+_{cc} \to  \Sigma^+ D^0 $&$\sin \left(\theta _c\right)$\\\hline
     $\Omega^+_{cc} \to  p D^0 $&$\Xi^+_{cc} \to  \Sigma^+ D^0 $&$-\sin ^2\left(\theta _c\right)$ &         $\Xi^{++}_{cc} \to  p D^+_s $&$\Xi^{++}_{cc} \to  \Sigma^+ D^+_s $&$-\sin \left(\theta _c\right)$\\\hline
      $\Omega^+_{cc} \to  n D^+ $&$\Xi^+_{cc} \to  \Xi^0D^+_s $&$-\sin ^2\left(\theta _c\right)$ &          $\Xi^{++}_{cc} \to  p D^+_s $&$\Xi^{++}_{cc} \to  p D^+ $&$-\sin \left(\theta _c\right)$\\\hline
      $\Xi^{++}_{cc} \to  \Sigma^+ D^+_s $&$\Xi^{++}_{cc} \to  \Sigma^+ D^+ $&$\sin \left(\theta _c\right)$ &          $\Omega^+_{cc} \to  p D^0 $&$\Xi^+_{cc} \to  p D^0 $&$-\sin \left(\theta _c\right)$\\\hline
       $\Xi^{++}_{cc} \to  p D^+ $&$\Xi^{++}_{cc} \to  \Sigma^+ D^+ $&$\sin \left(\theta _c\right)$ &           $\Omega^+_{cc} \to  p D^0 $&$\Omega^+_{cc} \to  \Sigma^+ D^0 $&$-\sin \left(\theta _c\right)$\\\hline
\end{tabular}
\end{table}

\begin{table}
\caption{ Relations for Cabibbo-allowed, singly Cabibbo suppressed and doubly Cabibbo suppressed processes in Tab.~\ref{tab:ccq_cqq6_qqbar}. }\label{tab:relations_cabibbo2}
\begin{tabular}{|c|c|c||c|c|c|c|c}\hline\hline
channel 1 & channel 2 & $r$ & channel 1 & channel 2 & $r$\\\hline
$\Xi^{++}_{cc} \to  \Sigma^{++}_c \pi^0 $&$\Xi^{++}_{cc} \to  \Sigma^{++}_c \eta$&$-\frac{1}{\sqrt{3}}$ &   $\Xi^+_{cc} \to  \Sigma^{++}_c K^- $&$\Omega^+_{cc} \to  \Sigma^{++}_c \pi^-$&$\csc ^2\left(\theta _c\right)$\\\hline
$\Xi^{++}_{cc} \to  \Sigma^{++}_c \pi^0 $&$\Xi^{++}_{cc} \to  \Sigma^{++}_c K^0 $&$\frac{\csc \left(\theta _c\right)}{\sqrt{2}}$ &       $\Xi^+_{cc} \to  \Sigma^{++}_c K^- $&$\Omega^+_{cc} \to  \Sigma^{++}_c K^- $&$\csc \left(\theta _c\right)$\\\hline
$\Xi^{++}_{cc} \to  \Sigma^{++}_c \pi^0 $&$\Xi^{++}_{cc} \to  \Sigma^{++}_c \bar{K}^0$&$\frac{\sin \left(\theta _c\right)}{\sqrt{2}}$ &       $\Xi^+_{cc} \to  \Sigma^+_c K^0 $&$\Omega^+_{cc} \to  \Xi^{'+}_c \bar{K}^0$&$\sin ^2\left(\theta _c\right)$\\\hline
 $\Xi^{++}_{cc} \to  \Sigma^{++}_c \eta$&$\Xi^{++}_{cc} \to  \Sigma^{++}_c K^0 $&$-\sqrt{\frac{3}{2}} \csc \left(\theta _c\right)$ &       $\Xi^+_{cc} \to  \Sigma^+_c \bar{K}^0$&$\Omega^+_{cc} \to  \Xi^{'+}_c K^0 $&$\csc ^2\left(\theta _c\right)$\\\hline
  $\Xi^{++}_{cc} \to  \Sigma^{++}_c \eta$&$\Xi^{++}_{cc} \to  \Sigma^{++}_c \bar{K}^0$&$-\sqrt{\frac{3}{2}} \sin \left(\theta _c\right)$ &        $\Xi^+_{cc} \to  \Xi^{'+}_c K^0 $&$\Omega^+_{cc} \to  \Sigma^+_c \bar{K}^0$&$-1$\\\hline
  $\Xi^{++}_{cc} \to  \Sigma^{++}_c K^0 $&$\Xi^{++}_{cc} \to  \Sigma^{++}_c \bar{K}^0$&$\sin ^2\left(\theta _c\right)$ &        $\Xi^+_{cc} \to  \Sigma^0_c \pi^+ $&$\Xi^+_{cc} \to  \Xi^{'0}_c \pi^+ $&$-\sqrt{2} \sin \left(\theta _c\right)$\\\hline
  $\Xi^{++}_{cc} \to  \Sigma^+_c \pi^+ $&$\Xi^{++}_{cc} \to  \Sigma^+_c K^+ $&$-\csc \left(\theta _c\right)$ &        $\Xi^+_{cc} \to  \Sigma^0_c \pi^+ $&$\Omega^+_{cc} \to  \Xi^{'0}_c K^+ $&$-\sqrt{2} \csc \left(\theta _c\right)$\\\hline
   $\Xi^{++}_{cc} \to  \Sigma^+_c \pi^+ $&$\Xi^{++}_{cc} \to  \Xi^{'+}_c \pi^+ $&$-\sin \left(\theta _c\right)$ &        $\Xi^+_{cc} \to  \Sigma^0_c \pi^+ $&$\Omega^+_{cc} \to  \Omega^0_{c} K^+ $&$-1$\\\hline
   $\Xi^{++}_{cc} \to  \Sigma^+_c \pi^+ $&$\Xi^{++}_{cc} \to  \Xi^{'+}_c K^+ $&$-1$ &         $\Xi^+_{cc} \to  \Sigma^0_c K^+ $&$\Omega^+_{cc} \to  \Omega^0_{c} \pi^+ $&$\sin ^2\left(\theta _c\right)$\\\hline
   $\Xi^{++}_{cc} \to  \Sigma^+_c K^+ $&$\Xi^{++}_{cc} \to  \Xi^{'+}_c \pi^+ $&$\sin ^2\left(\theta _c\right)$ &         $\Xi^+_{cc} \to  \Xi^{'0}_c \pi^+ $&$\Omega^+_{cc} \to  \Xi^{'0}_c K^+ $&$\csc ^2\left(\theta _c\right)$\\\hline
    $\Xi^{++}_{cc} \to  \Sigma^+_c K^+ $&$\Xi^{++}_{cc} \to  \Xi^{'+}_c K^+ $&$\sin \left(\theta _c\right)$ &          $\Xi^+_{cc} \to  \Xi^{'0}_c \pi^+ $&$\Omega^+_{cc} \to  \Omega^0_{c} K^+ $&$\frac{\csc \left(\theta _c\right)}{\sqrt{2}}$\\\hline
    $\Xi^{++}_{cc} \to  \Xi^{'+}_c \pi^+ $&$\Xi^{++}_{cc} \to  \Xi^{'+}_c K^+ $&$\csc \left(\theta _c\right)$ &          $\Xi^+_{cc} \to  \Xi^{'0}_c K^+ $&$\Omega^+_{cc} \to  \Xi^{'0}_c \pi^+ $&$-1$\\\hline
    $\Xi^+_{cc} \to  \Sigma^{++}_c \pi^-$&$\Xi^+_{cc} \to  \Sigma^{++}_c K^- $&$-\sin \left(\theta _c\right)$ &          $\Xi^+_{cc} \to  \Omega^0_{c} K^+ $&$\Omega^+_{cc} \to  \Sigma^0_c \pi^+ $&$\csc ^2\left(\theta _c\right)$\\\hline
     $\Xi^+_{cc} \to  \Sigma^{++}_c \pi^-$&$\Omega^+_{cc} \to  \Sigma^{++}_c \pi^-$&$-\csc \left(\theta _c\right)$ &           $\Omega^+_{cc} \to  \Sigma^{++}_c \pi^-$&$\Omega^+_{cc} \to  \Sigma^{++}_c K^- $&$\sin \left(\theta _c\right)$\\\hline
     $\Xi^+_{cc} \to  \Sigma^{++}_c \pi^-$&$\Omega^+_{cc} \to  \Sigma^{++}_c K^- $&$-1$ &            $\Omega^+_{cc} \to  \Xi^{'0}_c K^+ $&$\Omega^+_{cc} \to  \Omega^0_{c} K^+ $&$\frac{\sin \left(\theta _c\right)}{\sqrt{2}}$\\\hline
\end{tabular}
\end{table}

\begin{table}
\caption{Decay width relations for Cabibbo-allowed, singly Cabibbo suppressed and doubly Cabibbo suppressed processes in Tab.~\ref{tab:ccq_qqq10_cqbar}.}\label{tab:relations_cabibbo3}
\begin{tabular}{|c|c|c||c|c|c|c|c}\hline\hline
channel 1 & channel 2 & $r$ & channel 1 & channel 2 & $r$\\\hline
$\Xi^{++}_{cc} \to  \Delta^+ D^+ $&$\Xi^{++}_{cc} \to  \Delta^+ D^+_s $&$-\csc \left(\theta _c\right)$ &            $\Xi^+_{cc} \to  \Delta^0 D^+ $&$\Xi^+_{cc} \to  \Sigma^{'0} D^+ $&$-\sqrt{2} \sin \left(\theta _c\right)$\\\hline
 $\Xi^{++}_{cc} \to  \Delta^+ D^+ $&$\Xi^{++}_{cc} \to  \Sigma^{'+}D^+ $&$-\frac{1}{2} \sin \left(\theta _c\right)$ &             $\Xi^+_{cc} \to  \Delta^0 D^+ $&$\Omega^+_{cc} \to  \Sigma^{'0} D^+_s $&$-\sqrt{2} \csc \left(\theta _c\right)$\\\hline
  $\Xi^{++}_{cc} \to  \Delta^+ D^+ $&$\Xi^{++}_{cc} \to  \Sigma^{'+}D^+_s $&$-\frac{1}{2}$ &             $\Xi^+_{cc} \to  \Delta^0 D^+ $&$\Omega^+_{cc} \to  \Xi^{'0} D^+_s $&$-1$\\\hline
   $\Xi^{++}_{cc} \to  \Delta^+ D^+ $&$\Xi^+_{cc} \to  \Delta^0 D^+_s $&$-\frac{1}{2} \csc \left(\theta _c\right)$ &             $\Xi^+_{cc} \to  \Delta^0 D^+_s $&$\Omega^+_{cc} \to  \Xi^{'0} D^+ $&$\sin ^2\left(\theta _c\right)$\\\hline
    $\Xi^{++}_{cc} \to  \Delta^+ D^+ $&$\Omega^+_{cc} \to  \Xi^{'0} D^+ $&$-\frac{1}{2} \sin \left(\theta _c\right)$ &             $\Xi^+_{cc} \to  \Sigma^{'+}D^0 $&$\Xi^+_{cc} \to  \Xi^{'0} D^+_s $&$1$\\\hline
     $\Xi^{++}_{cc} \to  \Delta^+ D^+_s $&$\Xi^{++}_{cc} \to  \Sigma^{'+}D^+ $&$\frac{1}{2} \sin ^2\left(\theta _c\right)$ &              $\Xi^+_{cc} \to  \Sigma^{'+}D^0 $&$\Omega^+_{cc} \to  \Delta^+ D^0 $&$2 \csc ^2\left(\theta _c\right)$\\\hline
     $\Xi^{++}_{cc} \to  \Delta^+ D^+_s $&$\Xi^{++}_{cc} \to  \Sigma^{'+}D^+_s $&$\frac{\sin \left(\theta _c\right)}{2}$ &              $\Xi^+_{cc} \to  \Sigma^{'+}D^0 $&$\Omega^+_{cc} \to  \Delta^0 D^+ $&$\csc ^2\left(\theta _c\right)$\\\hline
      $\Xi^{++}_{cc} \to  \Delta^+ D^+_s $&$\Xi^+_{cc} \to  \Delta^0 D^+_s $&$\frac{1}{2}$ &              $\Xi^+_{cc} \to  \Sigma^{'+}D^0 $&$\Omega^+_{cc} \to  \Sigma^{'+}D^0 $&$\csc \left(\theta _c\right)$\\\hline
       $\Xi^{++}_{cc} \to  \Delta^+ D^+_s $&$\Omega^+_{cc} \to  \Xi^{'0} D^+ $&$\frac{1}{2} \sin ^2\left(\theta _c\right)$ &               $\Xi^+_{cc} \to  \Sigma^{'0} D^+ $&$\Omega^+_{cc} \to  \Sigma^{'0} D^+_s $&$\csc ^2\left(\theta _c\right)$\\\hline
       $\Xi^{++}_{cc} \to  \Sigma^{'+}D^+ $&$\Xi^{++}_{cc} \to  \Sigma^{'+}D^+_s $&$\csc \left(\theta _c\right)$ &                $\Xi^+_{cc} \to  \Sigma^{'0} D^+ $&$\Omega^+_{cc} \to  \Xi^{'0} D^+_s $&$\frac{\csc \left(\theta _c\right)}{\sqrt{2}}$\\\hline
       $\Xi^{++}_{cc} \to  \Sigma^{'+}D^+ $&$\Xi^+_{cc} \to  \Delta^0 D^+_s $&$\csc ^2\left(\theta _c\right)$ &                 $\Xi^+_{cc} \to  \Sigma^{'0} D^+_s $&$\Omega^+_{cc} \to  \Sigma^{'0} D^+ $&$-1$\\\hline
       $\Xi^{++}_{cc} \to  \Sigma^{'+}D^+ $&$\Omega^+_{cc} \to  \Xi^{'0} D^+ $&$1$ &                  $\Xi^+_{cc} \to  \Xi^{'0} D^+_s $&$\Omega^+_{cc} \to  \Delta^+ D^0 $&$2 \csc ^2\left(\theta _c\right)$\\\hline
        $\Xi^{++}_{cc} \to  \Sigma^{'+}D^+_s $&$\Xi^+_{cc} \to  \Delta^0 D^+_s $&$\csc \left(\theta _c\right)$ &                   $\Xi^+_{cc} \to  \Xi^{'0} D^+_s $&$\Omega^+_{cc} \to  \Delta^0 D^+ $&$\csc ^2\left(\theta _c\right)$\\\hline
         $\Xi^{++}_{cc} \to  \Sigma^{'+}D^+_s $&$\Omega^+_{cc} \to  \Xi^{'0} D^+ $&$\sin \left(\theta _c\right)$ &                    $\Xi^+_{cc} \to  \Xi^{'0} D^+_s $&$\Omega^+_{cc} \to  \Sigma^{'+}D^0 $&$\csc \left(\theta _c\right)$\\\hline
         $\Xi^+_{cc} \to  \Delta^+ D^0 $&$\Xi^+_{cc} \to  \Sigma^{'+}D^0 $&$-\frac{1}{2} \sin \left(\theta _c\right)$ &                     $\Omega^+_{cc} \to  \Delta^+ D^0 $&$\Omega^+_{cc} \to  \Delta^0 D^+ $&$\frac{1}{2}$\\\hline
         $\Xi^+_{cc} \to  \Delta^+ D^0 $&$\Xi^+_{cc} \to  \Xi^{'0} D^+_s $&$-\frac{1}{2} \sin \left(\theta _c\right)$ &                      $\Omega^+_{cc} \to  \Delta^+ D^0 $&$\Omega^+_{cc} \to  \Sigma^{'+}D^0 $&$\frac{\sin \left(\theta _c\right)}{2}$\\\hline
         $\Xi^+_{cc} \to  \Delta^+ D^0 $&$\Omega^+_{cc} \to  \Delta^+ D^0 $&$-\csc \left(\theta _c\right)$ &                       $\Omega^+_{cc} \to  \Delta^0 D^+ $&$\Omega^+_{cc} \to  \Sigma^{'+}D^0 $&$\sin \left(\theta _c\right)$\\\hline
          $\Xi^+_{cc} \to  \Delta^+ D^0 $&$\Omega^+_{cc} \to  \Delta^0 D^+ $&$-\frac{1}{2} \csc \left(\theta _c\right)$ &                       $\Omega^+_{cc} \to  \Sigma^{'0} D^+_s $&$\Omega^+_{cc} \to  \Xi^{'0} D^+_s $&$\frac{\sin \left(\theta _c\right)}{\sqrt{2}}$\\\hline
           $\Xi^+_{cc} \to  \Delta^+ D^0 $&$\Omega^+_{cc} \to  \Sigma^{'+}D^0 $&$-\frac{1}{2}$ &   &   &   \\\hline
\end{tabular}
\end{table}


\section{Non-Leptonic $\Xi_{bb}$ and $\Omega_{bb}$ decays}
\label{sec:bbq_nonleptonic}

For the bottom quark decay, there are generically  4 kinds of quark-level transitions:
\begin{eqnarray}
b\to c\bar c d/s, \;
b\to c \bar u d/s, \;
b\to u \bar c d/s, \;
b\to q_1 \bar q_2 q_3,
\end{eqnarray}
with $q_{1,2,3}$ being the light quarks.
Each of them will induce more than one types of decay modes at hadron level, which will be analyzed in order in the following.

\subsection{$b\to c\bar c d/s$}

\subsubsection{Decays into $J/\psi$ plus a bottom baryon }

\begin{figure}
\includegraphics[width=0.3\columnwidth]{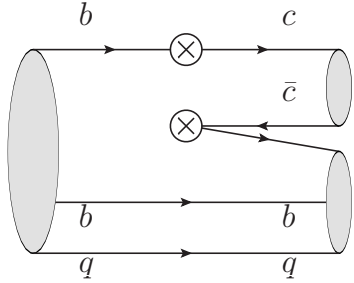}
\caption{Feynman diagrams for $\Xi_{bb}$ and $\Omega_{bb}$ decays into $J/\psi$ and a bottom baryon.  }
\label{bbq_bqqandjpsi}
\end{figure}

Such decays  have the same topology  with  semileptonic $b\to s\ell^+\ell^-$ decays.
The   transition operator $b\to c\bar cd/s$  can form an SU(3) triplet, leadings to the effective Hamiltonian:
\begin{eqnarray}
  H_{\rm{eff}}=   a_1 (T_{bb})^i (H_{  3})^j  (\overline T_{\bf{b\bar 3}})_{[ij]}~ J/\psi+ a_2 (T_{bb})^i  (H_{  3})^j  (\overline T_{\bf{b6}})_{\{ij\}}~J/\psi,
\end{eqnarray}
with $(H_{  3})_{2}=V_{cd}^*$ and $(H_{  3})_{3}=V_{cs}^*$. Feynman diagrams for these decays are given in Fig.~\ref{bbq_bqqandjpsi}.

Decay amplitudes are given in Tab.~\ref{tab:bbq_Jpsi}. Thus we have the following relations for decay widths:
\begin{eqnarray}
    \Gamma(\Xi_{bb}^{0}\to\Lambda_b^0J/\psi)&=& { }\Gamma(\Omega_{bb}^{-}\to\Xi_b^-J/\psi),\\
    \Gamma(\Xi_{bb}^{0}\to\Xi_b^0J/\psi)&=& { }\Gamma(\Xi_{bb}^{-}\to\Xi_b^-J/\psi),\\
    \Gamma(\Xi_{bb}^{0}\to\Sigma_{b}^{0}J/\psi)&=&{ }\Gamma(\Omega_{bb}^{-}\to\Xi_{b}^{\prime-}J/\psi) = \frac{1}{2}\Gamma(\Xi_{bb}^{-}\to\Sigma_{b}^{-}J/\psi),\\
    \Gamma(\Xi_{bb}^{0}\to\Xi_{b}^{\prime0}J/\psi)&=& { }\Gamma(\Xi_{bb}^{-}\to\Xi_{b}^{\prime-}J/\psi)=\frac{1}{2}\Gamma(\Omega_{bb}^{-}\to\Omega_{b}^{-}J/\psi).
     \end{eqnarray}

 \begin{table}
\caption{Doubly bottom baryons decays into a $J/\psi$ and a light baryon.}\label{tab:bbq_Jpsi}\begin{tabular}{|c|c|c|c|c|c|c|c}\hline\hline
channel & amplitude \\\hline
$\Xi_{bb}^{0}\to \Lambda_b^0J/\psi $ & $ a_1 V_{\text{cd}}^*$\\\hline
$\Xi_{bb}^{0}\to \Xi_b^0J/\psi $ & $ a_1 V_{\text{cs}}^*$\\\hline
$\Xi_{bb}^{-}\to \Xi_b^-J/\psi $ & $ a_1 V_{\text{cs}}^*$\\\hline
$\Omega_{bb}^{-}\to \Xi_b^-J/\psi $ & $ -a_1 V_{\text{cd}}^*$\\\hline
\hline$\Xi_{bb}^{0}\to \Sigma_{b}^{0}J/\psi $ & $ \frac{a_2 V_{\text{cd}}^*}{\sqrt{2}}$\\\hline
$\Xi_{bb}^{0}\to \Xi_{b}^{\prime0}J/\psi $ & $ \frac{a_2 V_{\text{cs}}^*}{\sqrt{2}}$\\\hline
$\Xi_{bb}^{-}\to \Sigma_{b}^{-}J/\psi $ & $ a_2 V_{\text{cd}}^*$\\\hline
$\Xi_{bb}^{-}\to \Xi_{b}^{\prime-}J/\psi $ & $ \frac{a_2 V_{\text{cs}}^*}{\sqrt{2}}$\\\hline
$\Omega_{bb}^{-}\to \Xi_{b}^{\prime-}J/\psi $ & $ \frac{a_2 V_{\text{cd}}^*}{\sqrt{2}}$\\\hline
$\Omega_{bb}^{-}\to \Omega_{b}^{-}J/\psi $ & $ a_2 V_{\text{cs}}^*$\\\hline
\hline
\end{tabular}
\end{table}

\subsubsection{Decays into a doubly heavy baryon $bcq$  plus a  anti-charmed  meson  }

\begin{figure}
\includegraphics[width=0.6\columnwidth]{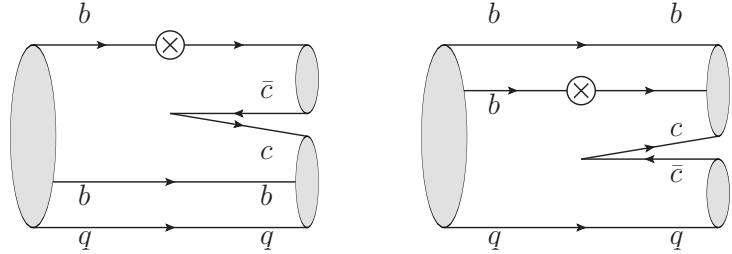}
\caption{Feynman diagrams for $\Xi_{bb}$ and $\Omega_{bb}$ decays into an doubly heavy baryon and an anti-charmed meson.  }
\label{bbq_doublyheavyandanticharmedmeson}
\end{figure}

The   $b\to c\bar cd/s$   transition can lead to another type of effective Hamiltonian:
\begin{eqnarray}
  H_{\rm{eff}}=   a_3  (T_{bb})^i  (H_{  3})^j  (\overline T_{bc})_i  D_j+ a_4 (T_{bb})^i  (H_{ 3})^j  (\overline T_{bc})_j D_i,
\end{eqnarray}
which corresponds to the decays into doubly heavy baryon $bcq$  plus a  anti-charmed  meson. Feynman diagrams for these decays are given in Fig.~\ref{bbq_doublyheavyandanticharmedmeson}.
Decay amplitudes are given in Tab.~\ref{tab:bbq_bcq_cbarq}. Thus we obtain the following relations for decay widths:
\begin{eqnarray}
    \Gamma(\Xi_{bb}^{0}\to\Xi_{bc}^{0}\overline D^0)= { }\Gamma(\Omega_{bb}^{-}\to\Xi_{bc}^{0}D^-_s),\nonumber\\
    \Gamma(\Xi_{bb}^{0}\to\Omega_{bc}^{0}\overline D^0)= { }\Gamma(\Xi_{bb}^{-}\to\Omega_{bc}^{0}D^-),\nonumber\\
     \Gamma(\Xi_{bb}^{0}\to\Xi_{bc}^{+}D^-)= { }\Gamma(\Omega_{bb}^{-}\to\Omega_{bc}^{0}D^-),\nonumber\\
     \Gamma(\Xi_{bb}^{0}\to\Xi_{bc}^{+}D^-_s)= { }\Gamma(\Xi_{bb}^{-}\to\Xi_{bc}^{0}D^-_s).
      \end{eqnarray}

  \begin{table}
\caption{Doubly bottom baryons decays into a $bcq$ and an anti-charmed meson.}\label{tab:bbq_bcq_cbarq}\begin{tabular}{|c|c|c|c|c|c|c|c}\hline\hline
channel & amplitude \\\hline
$\Xi_{bb}^{0}\to \Xi_{bc}^{+}  D^- $ & $ a_3 V_{\text{cd}}^*$\\\hline
$\Xi_{bb}^{0}\to \Xi_{bc}^{+}  D^-_s $ & $ a_3 V_{\text{cs}}^*$\\\hline
$\Xi_{bb}^{0}\to \Xi_{bc}^{0}  \overline D^0 $ & $ a_4 V_{\text{cd}}^*$\\\hline
$\Xi_{bb}^{0}\to \Omega_{bc}^{0}  \overline D^0 $ & $ a_4 V_{\text{cs}}^*$\\\hline
$\Xi_{bb}^{-}\to \Xi_{bc}^{0}  D^- $ & $ \left(a_3+a_4\right) V_{\text{cd}}^*$\\\hline
$\Xi_{bb}^{-}\to \Xi_{bc}^{0}  D^-_s $ & $ a_3 V_{\text{cs}}^*$\\\hline
$\Xi_{bb}^{-}\to \Omega_{bc}^{0}  D^- $ & $ a_4 V_{\text{cs}}^*$\\\hline
$\Omega_{bb}^{-}\to \Xi_{bc}^{0}  D^-_s $ & $ a_4 V_{\text{cd}}^*$\\\hline
$\Omega_{bb}^{-}\to \Omega_{bc}^{0}  D^- $ & $ a_3 V_{\text{cd}}^*$\\\hline
$\Omega_{bb}^{-}\to \Omega_{bc}^{0}  D^-_s $ & $ \left(a_3+a_4\right) V_{\text{cs}}^*$\\\hline
\hline
\end{tabular}
\end{table}

\subsection{$b\to c \bar u d/s$ transition}

\subsubsection{Decays into a doubly heavy baryon $bcq$ plus a light meson }

\begin{figure}
\includegraphics[width=0.9\columnwidth]{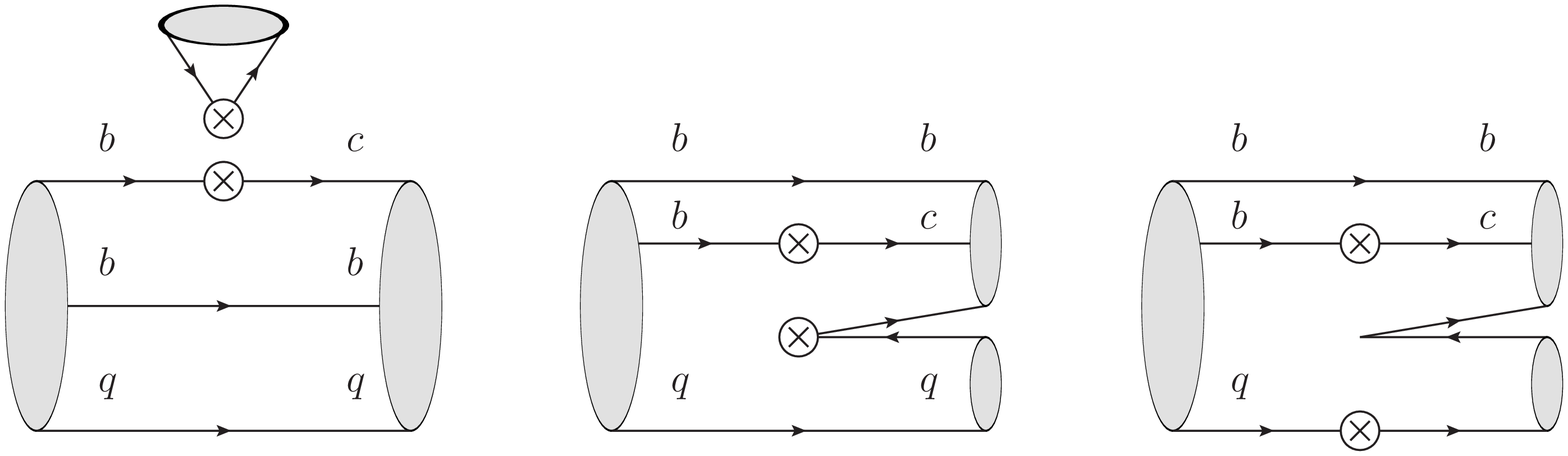}
\caption{Feynman diagrams for $\Xi_{bb}$ and $\Omega_{bb}$ decays into a doubly heavy baryon and a light meson.  }
\label{bbq_doublelyheavybaryonandmeson}
\end{figure}

The operator to produce a charm quark  from the $b$-quark decay, $\bar c b \bar q u$, is given by
\begin{eqnarray}
{\cal H}_{eff} &=& \frac{G_{F}}{\sqrt{2}}
     V_{cb} V_{uq}^{*} \big[
     C_{1}  O^{\bar cu}_{1}
  +  C_{2}  O^{\bar cu}_{2}\Big] +{\rm h.c.} .
\end{eqnarray}
The light quarks in this effective Hamiltonian form an octet with the nonzero entry
\begin{eqnarray}
(H_{{\bf8}})^2_1 =V_{ud}^*,
\end{eqnarray}
for   the $b\to c\bar ud$ transition, and   $(H_{{\bf8}})^3_1 =V_{us}^*$ for  the $b\to c\bar
us$ transition.
The hadron-level effective Hamiltonian is then given as
\begin{eqnarray}
  {\cal H}_{\rm{eff}}= a_5 (T_{bb})^i (\overline T_{bc})_i M^k_j (H_{{\bf8}})^j_k+ a_6 (T_{bb})^i (\overline T_{bc})_j M^k_i (H_{{\bf8}})^j_k+a_7 (T_{bb})^i (\overline T_{bc})_k M^k_j (H_{{\bf8}})^j_i.
\end{eqnarray}
Feynman diagrams for these decays are given in Fig.~\ref{bbq_doublelyheavybaryonandmeson}.
Decay amplitudes are given in Tab.~\ref{tab:bbq_bcq_qqbar}, which  leads to:
\begin{eqnarray}
\Gamma(\Xi_{bb}^{0}\to\Omega_{bc}^{0}\pi^0 )= \frac{1}{2}\Gamma(\Xi_{bb}^{-}\to\Omega_{bc}^{0}\pi^- ).
\end{eqnarray}

 \begin{table}
\caption{Doubly bottom baryons decays into a $bcq$ and a light meson.}\label{tab:bbq_bcq_qqbar}\begin{tabular}{|c  c|c  c|c|c|c|c|c|c}\hline\hline
channel & amplitude & channel & amplitude\\\hline
$\Xi_{bb}^{0}\to \Xi_{bc}^{+}  \pi^-  $ & $ \left(a_5+a_7\right) V_{\text{ud}}^*$ & $\Xi_{bb}^{0}\to \Omega_{bc}^{0}  \eta  $ & $ \frac{\left(a_6-2 a_7\right) V_{\text{us}}^*}{\sqrt{6}}$\\\hline
$\Xi_{bb}^{0}\to \Xi_{bc}^{+}  K^-  $ & $ \left(a_5+a_7\right) V_{\text{us}}^*$ & $\Xi_{bb}^{-}\to \Xi_{bc}^{0}  \pi^-  $ & $ \left(a_5+a_6\right) V_{\text{ud}}^*$\\\hline
$\Xi_{bb}^{0}\to \Xi_{bc}^{0}  \pi^0  $ & $ \frac{\left(a_6-a_7\right) V_{\text{ud}}^*}{\sqrt{2}}$ & $\Xi_{bb}^{-}\to \Xi_{bc}^{0}  K^-  $ & $ a_5 V_{\text{us}}^*$\\\hline
$\Xi_{bb}^{0}\to \Xi_{bc}^{0}  \overline K^0  $ & $ a_7 V_{\text{us}}^*$ & $\Xi_{bb}^{-}\to \Omega_{bc}^{0}  \pi^-  $ & $ a_6 V_{\text{us}}^*$\\\hline
$\Xi_{bb}^{0}\to \Xi_{bc}^{0}  \eta  $ & $ \frac{\left(a_6+a_7\right) V_{\text{ud}}^*}{\sqrt{6}}$ & $\Omega_{bb}^{-}\to \Xi_{bc}^{0}  K^-  $ & $ a_6 V_{\text{ud}}^*$\\\hline
$\Xi_{bb}^{0}\to \Omega_{bc}^{0}  \pi^0  $ & $ \frac{a_6 V_{\text{us}}^*}{\sqrt{2}}$ & $\Omega_{bb}^{-}\to \Omega_{bc}^{0}  \pi^-  $ & $ a_5 V_{\text{ud}}^*$\\\hline
$\Xi_{bb}^{0}\to \Omega_{bc}^{0}  K^0  $ & $ a_7 V_{\text{ud}}^*$ & $\Omega_{bb}^{-}\to \Omega_{bc}^{0}  K^-  $ & $ \left(a_5+a_6\right) V_{\text{us}}^*$\\\hline
\hline
\end{tabular}
\end{table}

\subsubsection{Decays into a bottom  baryon $bqq$ plus a charmed  meson  }

\begin{figure}
\includegraphics[width=0.6\columnwidth]{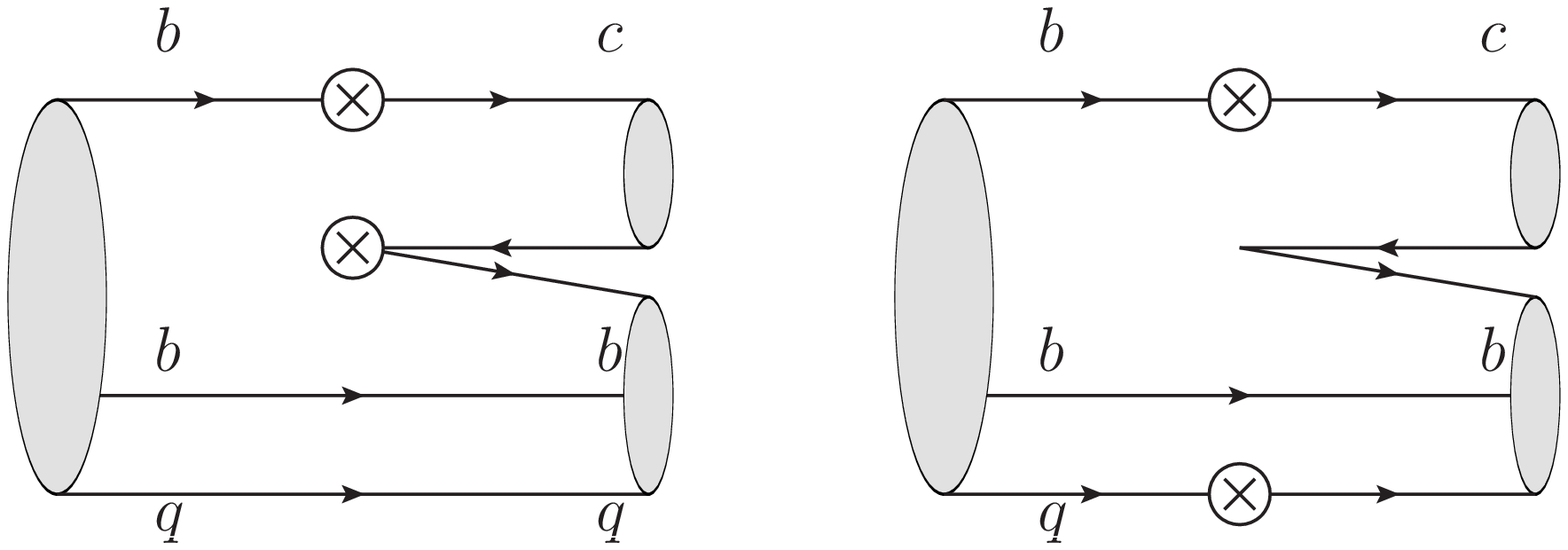}
\caption{Feynman diagrams for $\Xi_{bb}$ and $\Omega_{bb}$ decays into a bottom baryon and a charmed meson.  }
\label{bbq_bottombaryonandcharmedmeson}
\end{figure}

The effective Hamiltonian from the operator $\bar c b \bar q u$  gives
\begin{eqnarray}
  {\cal H}_{\rm{eff}}&=& a_8 (T_{bb})^i (\overline T_{b\bar 3})_{[ij]} \overline D^k (H_{{\bf8}})^j_k+ a_9 (T_{bb})^i (\overline T_{b\bar 3})_{[jk]} \overline D^k (H_{{\bf8}})^j_i  \nonumber\\
  &&+  a_{10} (T_{bb})^i (\overline T_{b6})_{\{ij\}} \overline D^k (H_{{\bf8}})^j_k+ a_{11} (T_{bb})^i (\overline T_{b6})_{\{jk\}} \overline D^k (H_{{\bf8}})^j_i.
\end{eqnarray}
Feynman diagrams for these decays are given in Fig.~\ref{bbq_bottombaryonandcharmedmeson}.
Results are given in Tab.~\ref{tab:bbq_bqq_cqbar}, thus we have the relations for decay amplitudes:
\begin{eqnarray}
    \Gamma(\Xi_{bb}^{0}\to\Sigma_{b}^{-}D^+)= 2\Gamma(\Xi_{bb}^{0}\to\Xi_{b}^{\prime-}D^+_s),\nonumber\\
    \Gamma(\Xi_{bb}^{0}\to\Xi_{b}^{\prime-}D^+)= \frac{1}{2}\Gamma(\Xi_{bb}^{0}\to\Omega_{b}^{-}D^+_s),\nonumber\\
      \Gamma(\Xi_{bb}^{-}\to\Sigma_{b}^{-}D^0)= 2\Gamma(\Omega_{bb}^{-}\to\Xi_{b}^{\prime-}D^0),\nonumber\\
      \Gamma(\Xi_{bb}^{-}\to\Xi_{b}^{\prime-}D^0)= \frac{1}{2}\Gamma(\Omega_{bb}^{-}\to\Omega_{b}^{-}D^0).
\end{eqnarray}

 \begin{table}
\caption{Doubly bottom baryons decays into a $bqq$ and a charmed meson.}\label{tab:bbq_bqq_cqbar}\begin{tabular}{|c  c|c  c|c|c|c|c|c|c}\hline\hline
channel & amplitude & channel & amplitude \\\hline
$\Xi_{bb}^{0}\to \Lambda_b^0  D^0 $ & $ \left(a_8-a_9\right) V_{\text{ud}}^*$ & $\Xi_{bb}^{0}\to \Sigma_{b}^{0}  D^0 $ & $ \frac{\left(a_{10}+a_{11}\right) V_{\text{ud}}^*}{\sqrt{2}}$\\\hline
$\Xi_{bb}^{0}\to \Xi_b^0  D^0 $ & $ \left(a_8-a_9\right) V_{\text{us}}^*$ & $\Xi_{bb}^{0}\to \Sigma_{b}^{-}  D^+ $ & $ a_{11} V_{\text{ud}}^*$\\\hline
$\Xi_{bb}^{0}\to \Xi_b^-  D^+ $ & $ -a_9 V_{\text{us}}^*$ & $\Xi_{bb}^{0}\to \Xi_{b}^{\prime0}  D^0 $ & $ \frac{\left(a_{10}+a_{11}\right) V_{\text{us}}^*}{\sqrt{2}}$\\\hline
$\Xi_{bb}^{0}\to \Xi_b^-  D^+_s $ & $ a_9 V_{\text{ud}}^*$ & $\Xi_{bb}^{0}\to \Xi_{b}^{\prime-}  D^+ $ & $ \frac{a_{11} V_{\text{us}}^*}{\sqrt{2}}$\\\hline
$\Xi_{bb}^{-}\to \Xi_b^-  D^0 $ & $ a_8 V_{\text{us}}^*$ & $\Xi_{bb}^{0}\to \Xi_{b}^{\prime-}  D^+_s $ & $ \frac{a_{11} V_{\text{ud}}^*}{\sqrt{2}}$\\\hline
$\Omega_{bb}^{-}\to \Xi_b^-  D^0 $ & $ -a_8 V_{\text{ud}}^*$ & $\Xi_{bb}^{0}\to \Omega_{b}^{-}  D^+_s $ & $ a_{11} V_{\text{us}}^*$\\\hline
&&$\Xi_{bb}^{-}\to \Sigma_{b}^{-}  D^0 $ & $ a_{10} V_{\text{ud}}^*$\\\hline
&&$\Xi_{bb}^{-}\to \Xi_{b}^{\prime-}  D^0 $ & $ \frac{a_{10} V_{\text{us}}^*}{\sqrt{2}}$\\\hline
&&$\Omega_{bb}^{-}\to \Xi_{b}^{\prime-}  D^0 $ & $ \frac{a_{10} V_{\text{ud}}^*}{\sqrt{2}}$\\\hline
&&$\Omega_{bb}^{-}\to \Omega_{b}^{-}  D^0 $ & $ a_{10} V_{\text{us}}^*$\\\hline
\hline
\end{tabular}
\end{table}

\subsection{$b\to u \bar c d/s$: Decays into a bottom  baryon $bqq$ plus an  anti-charmed  meson.  }

For the anti-charm production, the operator having the quark contents $(\bar ub)(\bar qc)$  is given by
\begin{eqnarray}
{\cal H}_{eff} &=& \frac{G_{F}}{\sqrt{2}}
     V_{ub} V_{cq}^{*} \big[
     C_{1}  O^{\bar uc}_{1}
  +  C_{2}  O^{\bar uc}_{2}\Big]+ {\rm h.c.}.
\end{eqnarray}
The two light anti-quarks form the ${\bf  \bar 3}$ and ${\bf  6}$ representations.
The anti-symmetric tensor $H_{\bar 3}''$ and the symmetric tensor
$H_{ 6}$ have nonzero components
\begin{eqnarray}
 (H_{\bar 3}'')^{13} =- (H_{\bar 3}'')^{31} =V_{cs}^*,\;\;\; (H_{\bar 6})^{13}=(H_{\bar 6})^{31} =V_{cs}^*,
\end{eqnarray}
for the $b\to u\bar cs$ transition. For the transition $b\to
u\bar cd$ one requests the interchange of $2\leftrightarrow 3$ in the
subscripts, and $V_{cs}$ replaced by $V_{cd}$.

\begin{figure}
\includegraphics[width=0.6\columnwidth]{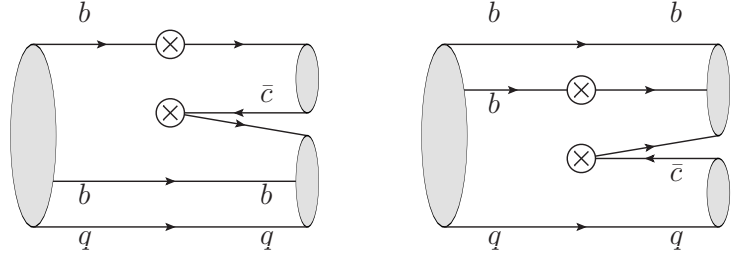}
\caption{Feynman diagrams for $\Xi_{bb}$ and $\Omega_{bb}$ decays into a bottom baryon and an anti-charmed meson.  }
\label{bbq_bqqandanticharmedmeson}
\end{figure}

The effective Hamiltonian is constructed as
\begin{eqnarray}
  {\cal H}_{\rm{eff}}&=& b_{1} (T_{bb})^i (\overline T_{b\bar 3})_{[ij]}   D_k (H_{\bar 3}'')^{jk}+b_{2} (T_{bb})^k (\overline T_{b\bar 3})_{[ij]}   D_k (H_{\bar  3}'')^{ij}+b_{3} (T_{bb})^i (\overline T_{b\bar 3})_{[ij]}   D_k (H_{  6}'')^{jk} \nonumber\\
  &&+  b_{4} (T_{bb})^i (\overline T_{b6})_{\{ij\}}   D_k (H_{ 6}'')^{jk}+b_{5} (T_{bb})^k (\overline T_{b6})_{\{ij\}}   D_k (H_{\bar 6}'')^{ij}+b_{6} (T_{bb})^i (\overline T_{b6})_{\{ij\}}   D_k (H_{\bar 3}'')^{jk}. \nonumber\\
\end{eqnarray}
Feynman diagrams for these decays are given in Fig.~\ref{bbq_bqqandanticharmedmeson}.
Decay amplitudes for different channels are given in Tab.~\ref{tab:bbq_bqq_qcbar},
thus we have the relations for decay amplitudes:
 \begin{eqnarray}
    \Gamma(\Xi_{bb}^{0}\to\Sigma_{b}^{+}D^-)= 2\Gamma(\Omega_{bb}^{-}\to\Xi_{b}^{\prime0}D^-),\\ \Gamma(\Xi_{bb}^{0}\to\Sigma_{b}^{+}D^-_s)= 2\Gamma(\Xi_{bb}^{-}\to\Sigma_{b}^{0}D^-_s),\\ \Gamma(\Xi_{bb}^{-}\to\Sigma_{b}^{-}\overline D^0)= 2\Gamma(\Omega_{bb}^{-}\to\Xi_{b}^{\prime-}\overline D^0),\\ \Gamma(\Xi_{bb}^{-}\to\Xi_{b}^{\prime-}\overline D^0)= \frac{1}{2}\Gamma(\Omega_{bb}^{-}\to\Omega_{b}^{-}\overline D^0).
    \end{eqnarray}

\begin{table}
\caption{Doubly bottom baryons decays into a $bqq$ and an anti-charmed meson.}\label{tab:bbq_bqq_qcbar}\begin{tabular}{|c  c|c  c|c|c|c|c|c|c}\hline\hline
channel & amplitude & channel & amplitude \\\hline
$\Xi_{bb}^{0}\to \Lambda_b^0  \overline D^0 $ & $ \left(-b_1+2 b_2+b_3\right) V_{\text{cd}}^*$ & $\Xi_{bb}^{0}\to \Sigma_{b}^{+}  D^- $ & $ \left(b_4+b_6\right) V_{\text{cd}}^*$\\\hline
$\Xi_{bb}^{0}\to \Xi_b^0  \overline D^0 $ & $ \left(-b_1+2 b_2+b_3\right) V_{\text{cs}}^*$ & $\Xi_{bb}^{0}\to \Sigma_{b}^{+}  D^-_s $ & $ \left(b_4+b_6\right) V_{\text{cs}}^*$\\\hline
$\Xi_{bb}^{-}\to \Lambda_b^0  D^- $ & $ -\left(b_1-2 b_2+b_3\right) V_{\text{cd}}^*$ & $\Xi_{bb}^{0}\to \Sigma_{b}^{0}  \overline D^0 $ & $ \frac{\left(b_4+2 b_5-b_6\right) V_{\text{cd}}^*}{\sqrt{2}}$\\\hline
$\Xi_{bb}^{-}\to \Lambda_b^0  D^-_s $ & $ -\left(b_1+b_3\right) V_{\text{cs}}^*$ & $\Xi_{bb}^{0}\to \Xi_{b}^{\prime0}  \overline D^0 $ & $ \frac{\left(b_4+2 b_5-b_6\right) V_{\text{cs}}^*}{\sqrt{2}}$\\\hline
$\Xi_{bb}^{-}\to \Xi_b^0  D^- $ & $ 2 b_2 V_{\text{cs}}^*$ & $\Xi_{bb}^{-}\to \Sigma_{b}^{0}  D^- $ & $ \frac{\left(b_4+2 b_5+b_6\right) V_{\text{cd}}^*}{\sqrt{2}}$\\\hline
$\Xi_{bb}^{-}\to \Xi_b^-  \overline D^0 $ & $ \left(b_3-b_1\right) V_{\text{cs}}^*$ & $\Xi_{bb}^{-}\to \Sigma_{b}^{0}  D^-_s $ & $ \frac{\left(b_4+b_6\right) V_{\text{cs}}^*}{\sqrt{2}}$\\\hline
$\Omega_{bb}^{-}\to \Lambda_b^0  D^-_s $ & $ 2 b_2 V_{\text{cd}}^*$ & $\Xi_{bb}^{-}\to \Sigma_{b}^{-}  \overline D^0 $ & $ \left(b_4-b_6\right) V_{\text{cd}}^*$\\\hline
$\Omega_{bb}^{-}\to \Xi_b^0  D^- $ & $ -\left(b_1+b_3\right) V_{\text{cd}}^*$ & $\Xi_{bb}^{-}\to \Xi_{b}^{\prime0}  D^- $ & $ \sqrt{2} b_5 V_{\text{cs}}^*$\\\hline
$\Omega_{bb}^{-}\to \Xi_b^0  D^-_s $ & $ -\left(b_1-2 b_2+b_3\right) V_{\text{cs}}^*$ & $\Xi_{bb}^{-}\to \Xi_{b}^{\prime-}  \overline D^0 $ & $ \frac{\left(b_4-b_6\right) V_{\text{cs}}^*}{\sqrt{2}}$\\\hline
$\Omega_{bb}^{-}\to \Xi_b^-  \overline D^0 $ & $ \left(b_1-b_3\right) V_{\text{cd}}^*$ & $\Omega_{bb}^{-}\to \Sigma_{b}^{0}  D^-_s $ & $ \sqrt{2} b_5 V_{\text{cd}}^*$\\\hline
&&$\Omega_{bb}^{-}\to \Xi_{b}^{\prime0}  D^- $ & $ \frac{\left(b_4+b_6\right) V_{\text{cd}}^*}{\sqrt{2}}$\\\hline
&&$\Omega_{bb}^{-}\to \Xi_{b}^{\prime0}  D^-_s $ & $ \frac{\left(b_4+2 b_5+b_6\right) V_{\text{cs}}^*}{\sqrt{2}}$\\\hline
&&$\Omega_{bb}^{-}\to \Xi_{b}^{\prime-}  \overline D^0 $ & $ \frac{\left(b_4-b_6\right) V_{\text{cd}}^*}{\sqrt{2}}$\\\hline
&&$\Omega_{bb}^{-}\to \Omega_{b}^{-}  \overline D^0 $ & $ \left(b_4-b_6\right) V_{\text{cs}}^*$\\\hline
\hline
\end{tabular}
\end{table}

As one can see, the $\Xi_{bb}$ can decay into both $\Xi_{b} D^0$ and $\Xi_{b} \overline D^0$.  The $D^0$ and $\overline D^0$ can form the CP eigenstate $D_{+}$ and $D_-$. Thus using the $\Xi_{bb}$ decays into the $\Xi_{b}D_\pm$, one may construct the interference between the $b\to c\bar us$ and $b\to u\bar cs$. The CKM angle $\gamma$ can then be extracted from measuring decay widths of these channels, as in the case of $B\to DK$~\cite{Gronau:1991dp,Gronau:1990ra,Dunietz:1991yd,Atwood:1996ci,hep-ph/0008090,Giri:2003ty},  $B\to DK^*_{0,2}$~\cite{Wang:2011zw,Kim:2013ria} and others. This is also similar for the $\Omega_{bb}\to \Omega^-D_\pm$ decays and the following $\Xi_{bc}\to\Xi_c D_\pm$ and $\Omega_{bc}\to \Omega^0 D_{\pm}$ modes.

\subsection{Charmless $b\to q_1 \bar q_2 q_3$ Decays}
\subsubsection{Decays into a bottom baryon and a light meson}

The  charmless $b\to q$ ($q=d,s$) transition is controlled by the weak Hamiltonian ${\cal H}_{eff}$:
 \begin{eqnarray}
 {\cal H}_{eff} &=& \frac{G_{F}}{\sqrt{2}}
     \bigg\{ V_{ub} V_{uq}^{*} \big[
     C_{1}  O^{\bar uu}_{1}
  +  C_{2}  O^{\bar uu}_{2}\Big]- V_{tb} V_{tq}^{*} \big[{\sum\limits_{i=3}^{10}} C_{i}  O_{i} \Big]\bigg\}+ \mbox{h.c.} ,
 \label{eq:hamiltonian}
\end{eqnarray}
where  $O_{i}$ is a four-quark operator or a moment type operator. At the hadron level,  penguin operators  behave as the ${\bf  3}$ representation while  tree operators   can
be decomposed in terms of a vector $H_{\bf 3}$, a traceless
tensor antisymmetric in upper indices, $H_{\bf\overline6}$, and a
traceless tensor symmetric in   upper indices,
$H_{\bf{15}}$.
For the $\Delta S=0 (b\to d)$decays, the non-zero components of the effective Hamiltonian are~\cite{Savage:1989ub,He:2000ys,Hsiao:2015iiu}:
\begin{eqnarray}
 (H_3)^2=1,\;\;\;(H_{\overline6})^{12}_1=-(H_{\overline6})^{21}_1=(H_{\overline6})^{23}_3=-(H_{\overline6})^{32}_3=1,\nonumber\\
 2(H_{15})^{12}_1= 2(H_{15})^{21}_1=-3(H_{15})^{22}_2=
 -6(H_{15})^{23}_3=-6(H_{15})^{32}_3=6,\label{eq:H3615_bb}
\end{eqnarray}
and  all other remaining entries are zero. For the $\Delta S=1(b\to s)$
decays the nonzero entries in the $H_{\bf{3}}$, $H_{\bf\overline6}$,
$H_{\bf{15}}$ are obtained from Eq.~\eqref{eq:H3615_bb}
with the exchange  $2\leftrightarrow 3$.

\begin{figure}
\includegraphics[width=0.8\columnwidth]{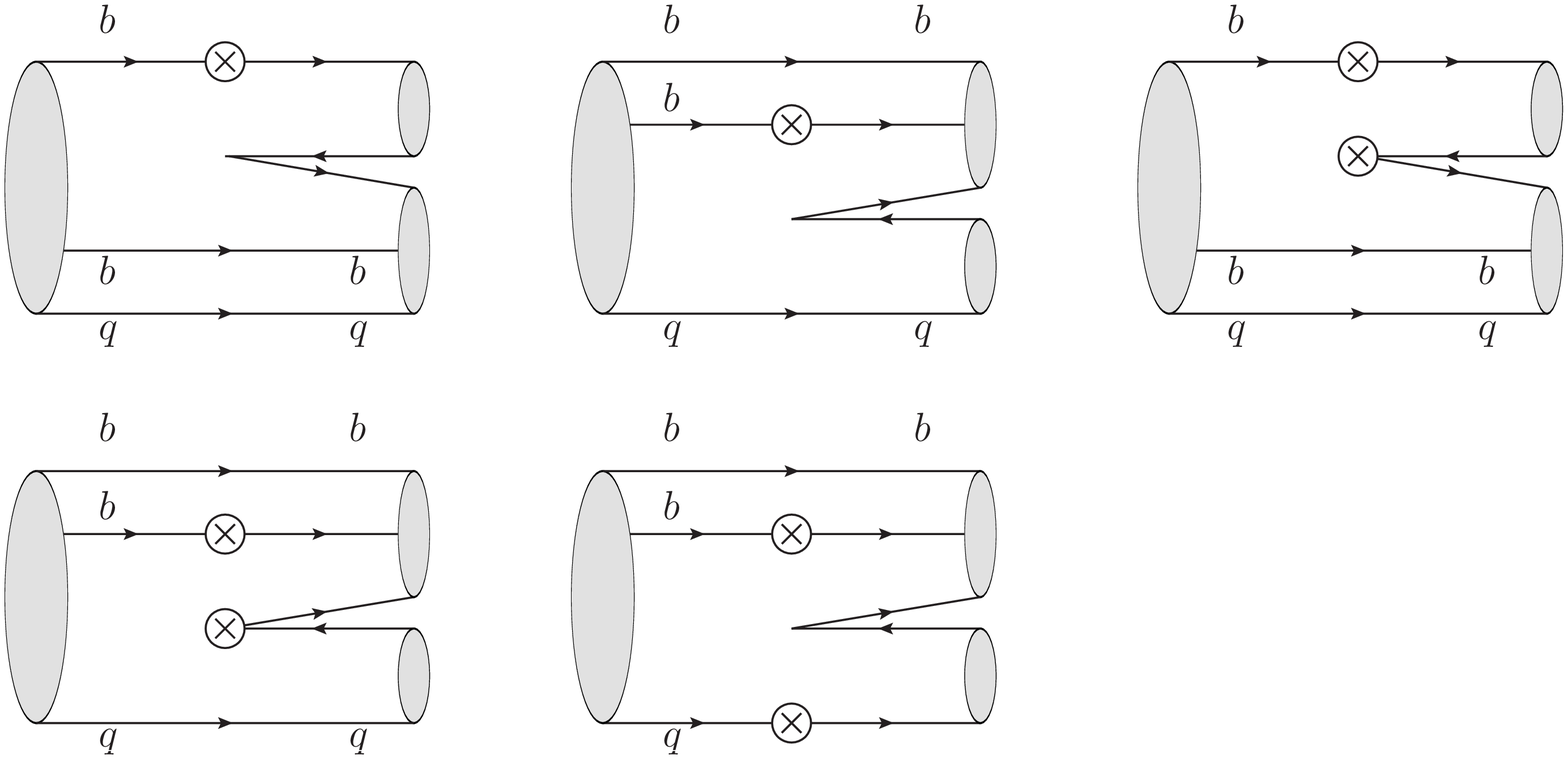}
\caption{Feynman diagrams for $\Xi_{bb}$ and $\Omega_{bb}$ decays into a bottom baryon and a light meson.  }
\label{bbq_bqqandlightmeson}
\end{figure}

The effective hadron-level Hamiltonian for decays into the bottom anti-triplet is constructed  as
\begin{eqnarray}
 {\cal H}_{eff}&=&c_1(T_{bb})^i  (\overline T_{b\bar 3})_{[ij]} M^{j}_{l}  (H_{3})^l +c_2(T_{bb})^i  (\overline T_{b\bar 3})_{[jl]} M^{j}_{i}  (H_{3})^l  \nonumber\\
 && + c_3(T_{bb})^i  (\overline T_{b\bar 3})_{[ij]} M^{k}_{l}  (H_{\bar 6})^{jl}_{k}+ c_4(T_{bb})^i  (\overline T_{b\bar 3})_{[jl]} M^{k}_{i}  (H_{\bar 6})^{jl}_{k}+ c_5(T_{bb})^i  (\overline T_{b\bar 3})_{[jk]} M^{k}_{l}  (H_{\bar 6})^{jl}_{i} \nonumber\\
 && +  c_6(T_{bb})^i  (\overline T_{b\bar 3})_{[ij]} M^{k}_{l}  (H_{15})^{jl}_{k}+ c_7(T_{bb})^i  (\overline T_{b\bar 3})_{[jk]} M^{k}_{l}  (H_{15})^{jl}_{i},
\end{eqnarray}
while for the sextet baryon, we have
\begin{eqnarray}
 {\cal H}_{eff}&=&c_8(T_{bb})^i  (\overline T_{b6})_{\{ij\}} M^{j}_{l}  (H_{3})^l +c_9(T_{bb})^i  (\overline T_{b6})_{\{jl\}} M^{j}_{i}  (H_{3})^l  \nonumber\\
 && + c_{10}(T_{bb})^i  (\overline T_{b6})_{\{ij\}} M^{k}_{l}  (H_{15})^{jl}_{k}+ c_{11}(T_{bb})^i  (\overline T_{b6})_{\{jl\}} M^{k}_{i}  (H_{15})^{jl}_{k}+ c_{12}(T_{bb})^i  (\overline T_{b6})_{\{jk\}} M^{k}_{l}  (H_{15})^{jl}_{i} \nonumber\\
 && +  c_{13}(T_{bb})^i  (\overline T_{b6})_{\{ij\}} M^{k}_{l}  (H_{\bar 6})^{jl}_{k}+ c_{14}(T_{bb})^i  (\overline T_{b6})_{\{jk\}} M^{k}_{l}  (H_{\bar 6})^{jl}_{i}.
\end{eqnarray}
Feynman diagrams for these decays are given in Fig.~\ref{bbq_bqqandlightmeson}. Decay amplitudes for different channels are given in Tab.~\ref{tab:bbq_bqq_qqbar_vd} and Tab.~\ref{tab:bbq_bqq_qqbar_vs} for $b\to d$ transition and $b\to s$ transition respectively.
Thus, it leads to the relations for decay widths:
\begin{eqnarray}
    \Gamma(\Xi_{bb}^{0}\to\Sigma_{b}^{-}\pi^+ )= 2\Gamma(\Xi_{bb}^{0}\to\Xi_{b}^{\prime-}K^+ ),\\
    \Gamma(\Xi_{bb}^{0}\to\Xi_{b}^{\prime-}\pi^+ )= \frac{1}{2}\Gamma(\Xi_{bb}^{0}\to\Omega_{b}^{-}K^+ ). \end{eqnarray}

 \begin{table}
\caption{Doubly bottom baryons decays into a $bqq$ and  a light meson induced by the charmless $b\to d$ transition.}\label{tab:bbq_bqq_qqbar_vd}\begin{tabular}{|c c|c c|c|c|c|c|c|c}\hline\hline
channel & amplitude & channel & amplitude\\\hline
$\Xi_{bb}^{0}\to \Lambda_b^0  \pi^0  $ & $ -\frac{c_1-c_2+c_3-2 c_4-5 c_6+6 c_7}{\sqrt{2}}$ & $\Xi_{bb}^{0}\to \Sigma_{b}^{+}  \pi^-  $ & $ c_8+3 c_{10}+3 c_{12}+c_{13}+c_{14}$\\\hline
$\Xi_{bb}^{0}\to \Lambda_b^0  \eta  $ & $ \frac{c_1+c_2-3 c_3+2 c_4+2 c_5+3 c_6}{\sqrt{6}}$ & $\Xi_{bb}^{0}\to \Sigma_{b}^{0}  \pi^0  $ & $ \frac{1}{2} \left(-c_8+c_9+5 c_{10}+6 c_{11}-c_{13}-2 c_{14}\right)$\\\hline
$\Xi_{bb}^{0}\to \Xi_b^0  K^0  $ & $ c_1-c_3+c_5-c_6+3 c_7$ & $\Xi_{bb}^{0}\to \Sigma_{b}^{0}  \eta  $ & $ \frac{c_8+c_9+3 c_{10}+6 c_{11}+6 c_{12}-3 c_{13}}{2 \sqrt{3}}$\\\hline
$\Xi_{bb}^{0}\to \Xi_b^-  K^+  $ & $ -c_2+2 c_4-c_5+3 c_7$ & $\Xi_{bb}^{0}\to \Sigma_{b}^{-}  \pi^+  $ & $ c_9-2 c_{11}+3 c_{12}-c_{14}$\\\hline
$\Xi_{bb}^{-}\to \Lambda_b^0  \pi^-  $ & $ -c_1+c_2-c_3+2 c_4-3 c_6+2 c_7$ & $\Xi_{bb}^{0}\to \Xi_{b}^{\prime0}  K^0  $ & $ \frac{c_8-c_{10}+3 c_{12}-c_{13}+c_{14}}{\sqrt{2}}$\\\hline
$\Xi_{bb}^{-}\to \Xi_b^-  K^0  $ & $ c_1-c_2-c_3+2 c_4-c_6-2 c_7$ & $\Xi_{bb}^{0}\to \Xi_{b}^{\prime-}  K^+  $ & $ \frac{c_9-2 c_{11}+3 c_{12}-c_{14}}{\sqrt{2}}$\\\hline
$\Omega_{bb}^{-}\to \Lambda_b^0  K^-  $ & $ c_2+2 c_4-c_5+c_7$ & $\Xi_{bb}^{-}\to \Sigma_{b}^{0}  \pi^-  $ & $ \frac{c_8+c_9+3 c_{10}+6 c_{11}-2 c_{12}+c_{13}}{\sqrt{2}}$\\\hline
$\Omega_{bb}^{-}\to \Xi_b^0  \pi^-  $ & $ -c_1-c_3+c_5-3 c_6+c_7$ & $\Xi_{bb}^{-}\to \Sigma_{b}^{-}  \pi^0  $ & $ -\frac{c_8+c_9-5 c_{10}-2 c_{11}-2 c_{12}+c_{13}}{\sqrt{2}}$\\\hline
$\Omega_{bb}^{-}\to \Xi_b^-  \pi^0  $ & $ \frac{c_1+c_3-c_5-5 c_6-c_7}{\sqrt{2}}$ & $\Xi_{bb}^{-}\to \Sigma_{b}^{-}  \eta  $ & $ \frac{c_8+c_9+3 c_{10}-2 c_{11}-2 c_{12}-3 c_{13}}{\sqrt{6}}$\\\hline
$\Omega_{bb}^{-}\to \Xi_b^-  \eta  $ & $ -\frac{c_1-2 c_2-3 c_3+4 c_4+c_5+3 c_6-3 c_7}{\sqrt{6}}$ & $\Xi_{bb}^{-}\to \Xi_{b}^{\prime-}  K^0  $ & $ \frac{c_8+c_9-c_{10}-2 c_{11}-2 c_{12}-c_{13}}{\sqrt{2}}$\\\hline
&&$\Omega_{bb}^{-}\to \Sigma_{b}^{0}  K^-  $ & $ \frac{c_9+6 c_{11}-c_{12}+c_{14}}{\sqrt{2}}$\\\hline
&&$\Omega_{bb}^{-}\to \Sigma_{b}^{-}  \overline K^0  $ & $ c_9-2 c_{11}-c_{12}+c_{14}$\\\hline
&&$\Omega_{bb}^{-}\to \Xi_{b}^{\prime0}  \pi^-  $ & $ \frac{c_8+3 c_{10}-c_{12}+c_{13}-c_{14}}{\sqrt{2}}$\\\hline
&&$\Omega_{bb}^{-}\to \Xi_{b}^{\prime-}  \pi^0  $ & $ \frac{1}{2} \left(-c_8+5 c_{10}+c_{12}-c_{13}+c_{14}\right)$\\\hline
&&$\Omega_{bb}^{-}\to \Xi_{b}^{\prime-}  \eta  $ & $ \frac{c_8-2 c_9+3 c_{10}+4 c_{11}+c_{12}-3 c_{13}-3 c_{14}}{2 \sqrt{3}}$\\\hline
&&$\Omega_{bb}^{-}\to \Omega_{b}^{-}  K^0  $ & $ c_8-c_{10}-c_{12}-c_{13}-c_{14}$\\\hline
\hline
\end{tabular}
\end{table}
  \begin{table}
\caption{Doubly bottom baryons decays into a $bqq$ and  a light meson induced by the charmless $b\to s$ transition.}\label{tab:bbq_bqq_qqbar_vs}\begin{tabular}{|c c|c c|c|c|c|c|c|c}\hline\hline
channel & amplitude & channel & amplitude\\\hline
$\Xi_{bb}^{0}\to \Lambda_b^0  \overline K^0  $ & $ c_1-c_3+c_5-c_6+3 c_7$ & $\Xi_{bb}^{0}\to \Sigma_{b}^{+}  K^-  $ & $ c_8+3 c_{10}+3 c_{12}+c_{13}+c_{14}$\\\hline
$\Xi_{bb}^{0}\to \Xi_b^0  \pi^0  $ & $ \frac{c_2-2 c_3+2 c_4+c_5+4 c_6-3 c_7}{\sqrt{2}}$ & $\Xi_{bb}^{0}\to \Sigma_{b}^{0}  \overline K^0  $ & $ \frac{c_8-c_{10}+3 c_{12}-c_{13}+c_{14}}{\sqrt{2}}$\\\hline
$\Xi_{bb}^{0}\to \Xi_b^0  \eta  $ & $ \frac{-2 c_1+c_2+2 c_4-c_5+6 c_6-9 c_7}{\sqrt{6}}$ & $\Xi_{bb}^{0}\to \Xi_{b}^{\prime0}  \pi^0  $ & $ \frac{1}{2} \left(c_9+4 c_{10}+6 c_{11}+3 c_{12}-2 c_{13}-c_{14}\right)$\\\hline
$\Xi_{bb}^{0}\to \Xi_b^-  \pi^+  $ & $ c_2-2 c_4+c_5-3 c_7$ & $\Xi_{bb}^{0}\to \Xi_{b}^{\prime0}  \eta  $ & $ \frac{-2 c_8+c_9+6 c_{10}+6 c_{11}-3 c_{12}-3 c_{14}}{2 \sqrt{3}}$\\\hline
$\Xi_{bb}^{-}\to \Lambda_b^0  K^-  $ & $ -c_1-c_3+c_5-3 c_6+c_7$ & $\Xi_{bb}^{0}\to \Xi_{b}^{\prime-}  \pi^+  $ & $ \frac{c_9-2 c_{11}+3 c_{12}-c_{14}}{\sqrt{2}}$\\\hline
$\Xi_{bb}^{-}\to \Xi_b^0  \pi^-  $ & $ c_2+2 c_4-c_5+c_7$ & $\Xi_{bb}^{0}\to \Omega_{b}^{-}  K^+  $ & $ c_9-2 c_{11}+3 c_{12}-c_{14}$\\\hline
$\Xi_{bb}^{-}\to \Xi_b^-  \pi^0  $ & $ -\frac{c_2+2 c_3-2 c_4-c_5-4 c_6+c_7}{\sqrt{2}}$ & $\Xi_{bb}^{-}\to \Sigma_{b}^{0}  K^-  $ & $ \frac{c_8+3 c_{10}-c_{12}+c_{13}-c_{14}}{\sqrt{2}}$\\\hline
$\Xi_{bb}^{-}\to \Xi_b^-  \eta  $ & $ \frac{-2 c_1+c_2-2 c_4+c_5+6 c_6+3 c_7}{\sqrt{6}}$ & $\Xi_{bb}^{-}\to \Sigma_{b}^{-}  \overline K^0  $ & $ c_8-c_{10}-c_{12}-c_{13}-c_{14}$\\\hline
$\Omega_{bb}^{-}\to \Xi_b^0  K^-  $ & $ -c_1+c_2-c_3+2 c_4-3 c_6+2 c_7$ & $\Xi_{bb}^{-}\to \Xi_{b}^{\prime0}  \pi^-  $ & $ \frac{c_9+6 c_{11}-c_{12}+c_{14}}{\sqrt{2}}$\\\hline
$\Omega_{bb}^{-}\to \Xi_b^-  \overline K^0  $ & $ -c_1+c_2+c_3-2 c_4+c_6+2 c_7$ & $\Xi_{bb}^{-}\to \Xi_{b}^{\prime-}  \pi^0  $ & $ \frac{1}{2} \left(-c_9+4 c_{10}+2 c_{11}+c_{12}-2 c_{13}-c_{14}\right)$\\\hline
&&$\Xi_{bb}^{-}\to \Xi_{b}^{\prime-}  \eta  $ & $ \frac{-2 c_8+c_9+6 c_{10}-2 c_{11}+c_{12}+3 c_{14}}{2 \sqrt{3}}$\\\hline
&&$\Xi_{bb}^{-}\to \Omega_{b}^{-}  K^0  $ & $ c_9-2 c_{11}-c_{12}+c_{14}$\\\hline
&&$\Omega_{bb}^{-}\to \Xi_{b}^{\prime0}  K^-  $ & $ \frac{c_8+c_9+3 c_{10}+6 c_{11}-2 c_{12}+c_{13}}{\sqrt{2}}$\\\hline
&&$\Omega_{bb}^{-}\to \Xi_{b}^{\prime-}  \overline K^0  $ & $ \frac{c_8+c_9-c_{10}-2 c_{11}-2 c_{12}-c_{13}}{\sqrt{2}}$\\\hline
&&$\Omega_{bb}^{-}\to \Omega_{b}^{-}  \pi^0  $ & $ \sqrt{2} \left(2 c_{10}-c_{13}\right)$\\\hline
&&$\Omega_{bb}^{-}\to \Omega_{b}^{-}  \eta  $ & $ -\sqrt{\frac{2}{3}} \left(c_8+c_9-3 c_{10}-2 c_{11}-2 c_{12}\right)$\\\hline
\hline
\end{tabular}
\end{table}



\subsubsection{ Decays into a bottom meson and a light baryon octet}

The effective Hamiltonian is given as
\begin{eqnarray}
 {\cal H}_{eff}&=&d_1(T_{bb})^i  \overline B ^j  \epsilon_{ijk} (T_8)^{k}_{l}  (H_{3})^l + d_2(T_{bb})^i   \overline B ^l  \epsilon_{ijk} (T_8)^{k}_{l}  (H_{3})^j+ d_3(T_{bb})^l   \overline B ^j  \epsilon_{ijk} (T_8)^{k}_{l}  (H_{3})^i
\nonumber\\
 && +d_4(T_{bb})^l  \overline B ^n  \epsilon_{ijk} (T_8)^{k}_{l}  (H_{6})^{ij}_n +d_5(T_{bb})^l  \overline B ^n  \epsilon_{ijk} (T_8)^{k}_{n}  (H_{6})^{ij}_l +d_6(T_{bb})^l  \overline B ^i\epsilon_{ijk} (T_8)^{k}_{n}  (H_{6})^{jn}_l   \nonumber\\
 &&+d_7(T_{bb})^i  \overline B^l\epsilon_{ijk} (T_8)^{k}_{n}  (H_{6})^{jn}_l+d_8(T_{bb})^l  \overline B ^i\epsilon_{ijk} (T_8)^{k}_{n}  (H_{15})^{jn}_l +d_9(T_{bb})^i  \overline B^l\epsilon_{ijk} (T_8)^{k}_{n}  (H_{15})^{jn}_l. \nonumber\\
\end{eqnarray}
Similarly, we find the reduced matrix elements $d_4,d_7$ and $d_5, d_6$ are not independent. So we use the following effective Hamiltonian:
\begin{eqnarray}
 {\cal H}_{eff}&=&d_1(T_{bb})^i  \overline B ^j  \epsilon_{ijk} (T_8)^{k}_{l}  (H_{3})^l + d_2(T_{bb})^i   \overline B ^l  \epsilon_{ijk} (T_8)^{k}_{l}  (H_{3})^j+ d_3(T_{bb})^l   \overline B ^j  \epsilon_{ijk} (T_8)^{k}_{l}  (H_{3})^i
\nonumber\\
 &&  +d_6(T_{bb})^l  \overline B ^i\epsilon_{ijk} (T_8)^{k}_{n}  (H_{6})^{jn}_l  +d_7(T_{bb})^i  \overline B^l\epsilon_{ijk} (T_8)^{k}_{n}  (H_{6})^{jn}_l \nonumber\\
 &&+d_8(T_{bb})^l  \overline B ^i\epsilon_{ijk} (T_8)^{k}_{n}  (H_{15})^{jn}_l +d_9(T_{bb})^i  \overline B^l\epsilon_{ijk} (T_8)^{k}_{n}  (H_{15})^{jn}_l.
\end{eqnarray}

\begin{figure}
\includegraphics[width=0.8\columnwidth]{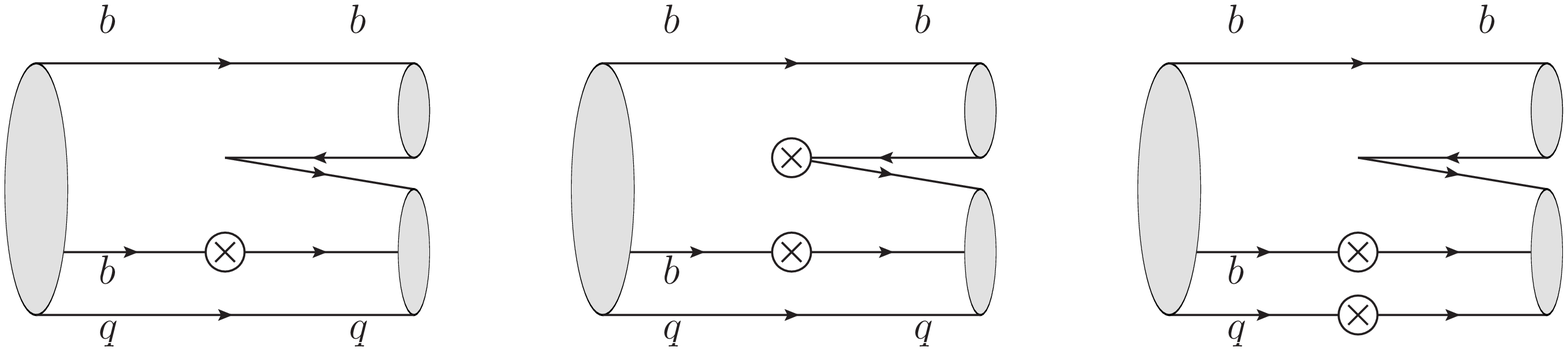}
\caption{Feynman diagrams for $\Xi_{bb}$ and $\Omega_{bb}$ decays into a bottom meson and a light baryon.  }
\label{bbq_qqq8andbottommeson}
\end{figure}

Feynman diagrams for these decays are given in Fig.~\ref{bbq_qqq8andbottommeson}. Decay amplitudes for different channels are given in Tab.~\ref{tab:bbq_qqq_bqbar_vd} and Tab.~\ref{tab:bbq_qqq_bqbar_vs} for $b\to d$ transition and $b\to s$ transition respectively.

  \begin{table}
\caption{Doubly bottom baryons decays into a bottom meson and  a light baryon induced by the charmless $b\to d$ transition.}\label{tab:bbq_qqq_bqbar_vd}\begin{tabular}{|c|c|c|c|c|c|c|c}\hline\hline
channel & amplitude \\\hline
$\Xi_{bb}^{0}\to \Lambda^0  \overline B^0_s $ & $ -\frac{d_1+2 d_2-d_3-2 d_6+d_7-3 d_9}{\sqrt{6}}$\\\hline
$\Xi_{bb}^{0}\to \Sigma^0  \overline B^0_s $ & $ \frac{d_1+d_3-d_7-6 d_8-d_9}{\sqrt{2}}$\\\hline
$\Xi_{bb}^{0}\to {p}  B^- $ & $ d_2-d_3-d_6-d_7+3 d_8+3 d_9$\\\hline
$\Xi_{bb}^{0}\to {n}  \overline B^0 $ & $ d_1+d_2-d_6-3 d_8-2 d_9$\\\hline
$\Xi_{bb}^{-}\to \Sigma^-  \overline B^0_s $ & $ d_1+d_3-d_7+2 d_8-d_9$\\\hline
$\Xi_{bb}^{-}\to {n}  B^- $ & $ -d_1-d_3-d_7-2 d_8-3 d_9$\\\hline
$\Omega_{bb}^{-}\to \Lambda^0  B^- $ & $ \frac{d_1-d_2+2 d_3-d_6+2 d_7+3 d_8}{\sqrt{6}}$\\\hline
$\Omega_{bb}^{-}\to \Sigma^0  B^- $ & $ -\frac{d_1+d_2+d_6+d_8+6 d_9}{\sqrt{2}}$\\\hline
$\Omega_{bb}^{-}\to \Sigma^-  \overline B^0 $ & $ -d_1-d_2-d_6-d_8+2 d_9$\\\hline
$\Omega_{bb}^{-}\to \Xi^-  \overline B^0_s $ & $ -d_2+d_3-d_6-d_7+d_8+d_9$\\\hline
\hline$\Xi_{bb}^{0}\to \Delta^{+}  B^- $ & $ \frac{f_3+6 \left(f_4+f_5\right)}{\sqrt{3}}$\\\hline
$\Xi_{bb}^{0}\to \Delta^{0}  \overline B^0 $ & $ \frac{f_3-2 f_4+6 f_5}{\sqrt{3}}$\\\hline
$\Xi_{bb}^{0}\to \Sigma^{\prime0}  \overline B^0_s $ & $ \frac{f_3-2 f_4+6 f_5}{\sqrt{6}}$\\\hline
$\Xi_{bb}^{-}\to \Delta^{0}  B^- $ & $ \frac{f_3+6 f_4-2 f_5}{\sqrt{3}}$\\\hline
$\Xi_{bb}^{-}\to \Delta^{-}  \overline B^0 $ & $ f_3-2 \left(f_4+f_5\right)$\\\hline
$\Xi_{bb}^{-}\to \Sigma^{\prime-}  \overline B^0_s $ & $ \frac{f_3-2 \left(f_4+f_5\right)}{\sqrt{3}}$\\\hline
$\Omega_{bb}^{-}\to \Sigma^{\prime0}  B^- $ & $ \frac{f_3+6 f_4-2 f_5}{\sqrt{6}}$\\\hline
$\Omega_{bb}^{-}\to \Sigma^{\prime-}  \overline B^0 $ & $ \frac{f_3-2 \left(f_4+f_5\right)}{\sqrt{3}}$\\\hline
$\Omega_{bb}^{-}\to \Xi^{\prime-}  \overline B^0_s $ & $ \frac{f_3-2 \left(f_4+f_5\right)}{\sqrt{3}}$\\\hline
\hline
\end{tabular}
\end{table}
    \begin{table}
\caption{Doubly bottom baryons decays into a bottom meson and  a light baryon induced by the charmless $b\to s$ transition.}\label{tab:bbq_qqq_bqbar_vs}\begin{tabular}{|c|c|c|c|c|c|c|c}\hline\hline
channel & amplitude \\\hline
$\Xi_{bb}^{0}\to \Lambda^0  \overline B^0 $ & $ \frac{-2 d_1-d_2-d_3+d_6+d_7+9 d_8+3 d_9}{\sqrt{6}}$\\\hline
$\Xi_{bb}^{0}\to \Sigma^+  B^- $ & $ -d_2+d_3+d_6+d_7-3 d_8-3 d_9$\\\hline
$\Xi_{bb}^{0}\to \Sigma^0  \overline B^0 $ & $ \frac{d_2-d_3-d_6+d_7+3 d_8-d_9}{\sqrt{2}}$\\\hline
$\Xi_{bb}^{0}\to \Xi^0  \overline B^0_s $ & $ -d_1-d_2+d_6+3 d_8+2 d_9$\\\hline
$\Xi_{bb}^{-}\to \Lambda^0  B^- $ & $ \frac{2 d_1+d_2+d_3+d_6+d_7+3 d_8+9 d_9}{\sqrt{6}}$\\\hline
$\Xi_{bb}^{-}\to \Sigma^0  B^- $ & $ \frac{d_2-d_3+d_6-d_7-d_8+3 d_9}{\sqrt{2}}$\\\hline
$\Xi_{bb}^{-}\to \Sigma^-  \overline B^0 $ & $ d_2-d_3+d_6+d_7-d_8-d_9$\\\hline
$\Xi_{bb}^{-}\to \Xi^-  \overline B^0_s $ & $ d_1+d_2+d_6+d_8-2 d_9$\\\hline
$\Omega_{bb}^{-}\to \Xi^-  \overline B^0 $ & $ -d_1-d_3+d_7-2 d_8+d_9$\\\hline
$\Omega_{bb}^{-}\to \Xi^0  B^- $ & $ d_1+d_3+d_7+2 d_8+3 d_9$\\\hline
\hline$\Xi_{bb}^{0}\to \Sigma^{\prime+}  B^- $ & $ \frac{f_3+6 \left(f_4+f_5\right)}{\sqrt{3}}$\\\hline
$\Xi_{bb}^{0}\to \Sigma^{\prime0}  \overline B^0 $ & $ \frac{f_3-2 f_4+6 f_5}{\sqrt{6}}$\\\hline
$\Xi_{bb}^{0}\to \Xi^{\prime0}  \overline B^0_s $ & $ \frac{f_3-2 f_4+6 f_5}{\sqrt{3}}$\\\hline
$\Xi_{bb}^{-}\to \Sigma^{\prime0}  B^- $ & $ \frac{f_3+6 f_4-2 f_5}{\sqrt{6}}$\\\hline
$\Xi_{bb}^{-}\to \Sigma^{\prime-}  \overline B^0 $ & $ \frac{f_3-2 \left(f_4+f_5\right)}{\sqrt{3}}$\\\hline
$\Xi_{bb}^{-}\to \Xi^{\prime-}  \overline B^0_s $ & $ \frac{f_3-2 \left(f_4+f_5\right)}{\sqrt{3}}$\\\hline
$\Omega_{bb}^{-}\to \Xi^{\prime0}  B^- $ & $ \frac{f_3+6 f_4-2 f_5}{\sqrt{3}}$\\\hline
$\Omega_{bb}^{-}\to \Xi^{\prime-}  \overline B^0 $ & $ \frac{f_3-2 \left(f_4+f_5\right)}{\sqrt{3}}$\\\hline
$\Omega_{bb}^{-}\to \Omega^-  \overline B^0_s $ & $ f_3-2 \left(f_4+f_5\right)$\\\hline
\hline
\end{tabular}
\end{table}


\subsubsection{ Decays into a bottom meson and a light baryon decuplet}

The effective Hamiltonian is given as
\begin{eqnarray}
 {\cal H}_{eff}&=& f_3(T_{bb})^l   \overline B ^j (T_{10})_{ijl} (H_{3})^i +f_4(T_{bb})^l  \overline B ^n (T_{10})_{ijl}  (H_{15})^{ij}_n +f_5(T_{bb})^l  \overline B ^n  (T_{10})_{ijn}  (H_{15})^{ij}_l.
\end{eqnarray}
Feynman diagrams for these decays are same as Fig.~\ref{bbq_qqq8andbottommeson}. Decay amplitudes for different channels are given in Tab.~\ref{tab:bbq_qqq_bqbar_vd} and Tab.~\ref{tab:bbq_qqq_bqbar_vs} for $b\to d$ transition and $b\to s$ transition respectively.

We summarize the relations for decay widths for $\Xi_{bb}$ and $\Omega_{bb}$ decay into a bottom meson and a light baryon,
   \begin{eqnarray}
    \Gamma(\Xi_{bb}^{0}\to\Delta^{0}\overline B^0)&=& 2\Gamma(\Xi_{bb}^{0}\to\Sigma^{\prime0}\overline B^0_s),\\
     \Gamma(\Xi_{bb}^{-}\to\Delta^{-}\overline B^0)&=& 3\Gamma(\Xi_{bb}^{-}\to\Sigma^{\prime-}\overline B^0_s)= 3\Gamma(\Omega_{bb}^{-}\to\Xi^{\prime-}\overline B^0_s)= 3\Gamma(\Omega_{bb}^{-}\to\Sigma^{\prime-}\overline B^0),\\
      \Gamma(\Xi_{bb}^{-}\to\Delta^{0}B^-)&=& 2\Gamma(\Omega_{bb}^{-}\to\Sigma^{\prime0}B^-),\\
    \Gamma(\Xi_{bb}^{0}\to\Sigma^{\prime0}\overline B^0)&=& \frac{1}{2}\Gamma(\Xi_{bb}^{0}\to\Xi^{\prime0}\overline B^0_s),\\
    \Gamma(\Xi_{bb}^{-}\to\Sigma^{\prime-}\overline B^0)&=& { }\Gamma(\Xi_{bb}^{-}\to\Xi^{\prime-}\overline B^0_s)= \Gamma(\Omega_{bb}^{-}\to\Xi^{\prime-}\overline B^0)= \frac{1}{3}\Gamma(\Omega_{bb}^{-}\to\Omega^{-}\overline B^0_s),\\
     \Gamma(\Xi_{bb}^{-}\to\Sigma^{\prime0}B^-)&=& \frac{1}{2}\Gamma(\Omega_{bb}^{-}\to\Xi^{\prime0}B^-).
      \end{eqnarray}

\subsubsection{U-spin for $\Xi_{bb}$ and $\Omega_{bb}$ decays }

For $\Xi_{bb},\Omega_{bb}$ decays induced by the $b\to q_1q_2q_3$, there are two amplitudes with different CKM factors. We consider the connected decays with the decay amplitudes
\begin{eqnarray}
  A(\Delta S=0)&=& r\left( V_{ub}V^*_{ud}A_{\Xi_{bb},\Omega_{bb}}^T+ V_{tb}V_{td}^*A_{\Xi_{bb},\Omega_{bb}}^P  \right), \nonumber\\
   A(\Delta S=1)&=&  V_{ub}V^*_{us}A_{\Xi_{bb},\Omega_{bb}}^T+ V_{tb}V_{ts}^*A_{\Xi_{bb},\Omega_{bb}}^P .
\end{eqnarray}
As pointed out in Refs.~\cite{Deshpande:1994ii,He:1998rq,Gronau:2000zy}, there exists a relation for the CP violating quantity $\Delta=\Gamma-\bar \Gamma$. The relation about decay widths $\Gamma(\Delta S=i)$ and CP asymmetry $A_{CP}(\Delta S=i)$ is
\begin{eqnarray}\label{cprelation}
  \frac{A_{CP}(\Delta S=0)}{A_{CP}(\Delta S=1)}= -r^2 \frac{\Gamma(\Delta S=1)}{\Gamma(\Delta S=0)}.
\end{eqnarray}

In Tab.\ref{cprelationpairslightmeson} and Tab.\ref{cprelationpairsbottommeson}, we collect the $\Xi_{bb},\Omega_{bb}$ decay pairs related by U-spin. The CP asymmetries and decay widths for these pairs satisfy relation in Eq.~\eqref{cprelation}. The experiment data in future  is important to test flavor SU(3) symmetry and also CKM mechanism for CP violation.

\begin{table}[!h]
\caption{U-spin relations for $\Xi_{bb},\Omega_{bb}$ decays into a bottom baryon and a light meson. Results in the ``channel 1" are for $b\to d$ processes and the ones in the ``channel 2" are for $b\to s$ processes. r denotes the ratio of the two amplitudes.  }\label{cprelationpairslightmeson}\begin{tabular}{|c|c|c||c|c|c|c|c}\hline\hline
channel 1 & channel 2 & $r$ & channel 1 & channel 2 & $r$ \\\hline
$\Xi^0_{bb} \to  \Lambda^0_b \bar{K}^0$&$\Xi^0_{bb} \to  \Xi^0_b K^0 $&$1$ &     $\Xi^0_{bb} \to  \Xi^{'-}_b \pi^+ $&$\Xi^0_{bb} \to  \Xi^{'-}_b K^+ $&$1$\\\hline
 $\Xi^0_{bb} \to  \Xi^-_b \pi^+ $&$\Xi^0_{bb} \to  \Xi^-_b K^+ $&$-1$ &      $\Xi^0_{bb} \to  \Omega^-_{b} K^+ $&$\Xi^0_{bb} \to  \Sigma^-_b \pi^+ $&$1$\\\hline
 $\Xi^-_{bb} \to  \Lambda^0_b K^- $&$\Omega^-_{bb} \to  \Xi^0_b \pi^-$&$1$ &      $\Xi^0_{bb} \to  \Omega^-_{b} K^+ $&$\Xi^0_{bb} \to  \Xi^{'-}_b K^+ $&$\sqrt{2}$\\\hline
  $\Xi^-_{bb} \to  \Xi^0_b \pi^-$&$\Omega^-_{bb} \to  \Lambda^0_b K^- $&$1$ &       $\Xi^-_{bb} \to  \Sigma^0_b K^- $&$\Omega^-_{bb} \to  \Xi^{'0}_b \pi^-$&$1$\\\hline
  $\Omega^-_{bb} \to  \Xi^0_b K^- $&$\Xi^-_{bb} \to  \Lambda^0_b \pi^-$&$1$ &        $\Xi^-_{bb} \to  \Xi^{'0}_b \pi^-$&$\Omega^-_{bb} \to  \Sigma^0_b K^- $&$1$\\\hline
  $\Omega^-_{bb} \to  \Xi^-_b \bar{K}^0$&$\Xi^-_{bb} \to  \Xi^-_b K^0 $&$-1$ &        $\Xi^-_{bb} \to  \Sigma^-_b \bar{K}^0$&$\Omega^-_{bb} \to  \Omega^-_{b} K^0 $&$1$\\\hline
   $\Xi^0_{bb} \to  \Sigma^+_b K^- $&$\Xi^0_{bb} \to  \Sigma^+_b \pi^-$&$1$ &        $\Xi^-_{bb} \to  \Omega^-_{b} K^0 $&$\Omega^-_{bb} \to  \Sigma^-_b \bar{K}^0$&$1$\\\hline
   $\Xi^0_{bb} \to  \Sigma^0_b \bar{K}^0$&$\Xi^0_{bb} \to  \Xi^{'0}_b K^0 $&$1$ &       $\Omega^-_{bb} \to  \Xi^{'0}_b K^- $&$\Xi^-_{bb} \to  \Sigma^0_b \pi^-$&$1$\\\hline
    $\Xi^0_{bb} \to  \Xi^{'-}_b \pi^+ $&$\Xi^0_{bb} \to  \Sigma^-_b \pi^+ $&$\frac{1}{\sqrt{2}}$ &          $\Omega^-_{bb} \to  \Xi^{'-}_b \bar{K}^0$&$\Xi^-_{bb} \to  \Xi^{'-}_b K^0 $&$1$\\\hline
\end{tabular}
\end{table}

\begin{table}[!h]
\caption{U-spin relations for $\Xi_{bb},\Omega_{bb}$ decays into a bottom meson and a light baryon. Results in the ``channel 1" are for $b\to d$ processes and the ones in the ``channel 2" are for $b\to s$ processes. $r$ denotes the ratio of the two amplitudes.  }\label{cprelationpairsbottommeson}\begin{tabular}{|c|c|c||c|c|c|c|c}\hline\hline
channel 1 & channel 2 & $r$ & channel 1 & channel 2 & $r$ \\\hline
$\Xi^0_{bb} \to  p B^- $&$\Xi^0_{bb} \to  \Sigma^+ B^- $&$-1$ &         $\Xi^-_{bb} \to  \Delta^0 B^- $&$\Omega^-_{bb} \to  \Xi^{\prime0} B^- $&$1$\\\hline
 $\Xi^0_{bb} \to  n \bar{B}^0 $&$\Xi^0_{bb} \to  \Xi^0\bar{B}^0_s $&$-1$ &         $\Xi^-_{bb} \to  \Sigma^{\prime-} \bar{B}^0_s $&$\Xi^-_{bb} \to  \Sigma^{\prime-} \bar{B}^0 $&$1$\\\hline
  $\Xi^-_{bb} \to  \Sigma^-\bar{B}^0_s $&$\Omega^-_{bb} \to  \Xi^- \bar{B}^0 $&$-1$ &         $\Xi^-_{bb} \to  \Sigma^{\prime-} \bar{B}^0_s $&$\Xi^-_{bb} \to  \Xi^{\prime-} \bar{B}^0_s $&$1$\\\hline
   $\Xi^-_{bb} \to  n B^- $&$\Omega^-_{bb} \to  \Xi^0B^- $&$-1$ &          $\Xi^-_{bb} \to  \Sigma^{\prime-} \bar{B}^0_s $&$\Omega^-_{bb} \to  \Xi^{\prime-} \bar{B}^0 $&$1$\\\hline
    $\Omega^-_{bb} \to  \Sigma^-\bar{B}^0 $&$\Xi^-_{bb} \to  \Xi^- \bar{B}^0_s $&$-1$ &           $\Xi^-_{bb} \to  \Sigma^{\prime-} \bar{B}^0_s $&$\Omega^-_{bb} \to  \Omega^{-}\bar{B}^0_s $&$\frac{1}{\sqrt{3}}$\\\hline
     $\Omega^-_{bb} \to  \Xi^- \bar{B}^0_s $&$\Xi^-_{bb} \to  \Sigma^-\bar{B}^0 $&$-1$ &            $\Omega^-_{bb} \to  \Sigma^{\prime-} \bar{B}^0 $&$\Xi^-_{bb} \to  \Sigma^{\prime-} \bar{B}^0 $&$1$\\\hline
     $\Xi^0_{bb} \to  \Delta^0 \bar{B}^0 $&$\Xi^0_{bb} \to  \Sigma^{\prime0} \bar{B}^0 $&$\sqrt{2}$ &             $\Omega^-_{bb} \to  \Sigma^{\prime-} \bar{B}^0 $&$\Xi^-_{bb} \to  \Xi^{\prime-} \bar{B}^0_s $&$1$\\\hline
     $\Xi^0_{bb} \to  \Delta^0 \bar{B}^0 $&$\Xi^0_{bb} \to  \Xi^{\prime0} \bar{B}^0_s $&$1$ &              $\Omega^-_{bb} \to  \Sigma^{\prime-} \bar{B}^0 $&$\Omega^-_{bb} \to  \Xi^{\prime-} \bar{B}^0 $&$1$\\\hline
     $\Xi^0_{bb} \to  \Sigma^{\prime0} \bar{B}^0_s $&$\Xi^0_{bb} \to  \Sigma^{\prime0} \bar{B}^0 $&$1$ &               $\Omega^-_{bb} \to  \Sigma^{\prime-} \bar{B}^0 $&$\Omega^-_{bb} \to  \Omega^{-}\bar{B}^0_s $&$\frac{1}{\sqrt{3}}$\\\hline
      $\Xi^0_{bb} \to  \Sigma^{\prime0} \bar{B}^0_s $&$\Xi^0_{bb} \to  \Xi^{\prime0} \bar{B}^0_s $&$\frac{1}{\sqrt{2}}$ &               $\Omega^-_{bb} \to  \Sigma^{\prime0} B^- $&$\Xi^-_{bb} \to  \Sigma^{\prime0} B^- $&$1$\\\hline
      $\Xi^-_{bb} \to  \Delta^- \bar{B}^0 $&$\Xi^-_{bb} \to  \Sigma^{\prime-} \bar{B}^0 $&$\sqrt{3}$ &               $\Omega^-_{bb} \to  \Sigma^{\prime0} B^- $&$\Omega^-_{bb} \to  \Xi^{\prime0} B^- $&$\frac{1}{\sqrt{2}}$\\\hline
       $\Xi^-_{bb} \to  \Delta^- \bar{B}^0 $&$\Xi^-_{bb} \to  \Xi^{\prime-} \bar{B}^0_s $&$\sqrt{3}$ &                $\Omega^-_{bb} \to  \Xi^{\prime-} \bar{B}^0_s $&$\Xi^-_{bb} \to  \Sigma^{\prime-} \bar{B}^0 $&$1$\\\hline
       $\Xi^-_{bb} \to  \Delta^- \bar{B}^0 $&$\Omega^-_{bb} \to  \Xi^{\prime-} \bar{B}^0 $&$\sqrt{3}$ &                 $\Omega^-_{bb} \to  \Xi^{\prime-} \bar{B}^0_s $&$\Xi^-_{bb} \to  \Xi^{\prime-} \bar{B}^0_s $&$1$\\\hline
        $\Xi^-_{bb} \to  \Delta^- \bar{B}^0 $&$\Omega^-_{bb} \to  \Omega^{-}\bar{B}^0_s $&$1$ &                 $\Omega^-_{bb} \to  \Xi^{\prime-} \bar{B}^0_s $&$\Omega^-_{bb} \to  \Xi^{\prime-} \bar{B}^0 $&$1$\\\hline
        $\Xi^-_{bb} \to  \Delta^0 B^- $&$\Xi^-_{bb} \to  \Sigma^{\prime0} B^- $&$\sqrt{2}$ &                 $\Omega^-_{bb} \to  \Xi^{\prime-} \bar{B}^0_s $&$\Omega^-_{bb} \to  \Omega^{-}\bar{B}^0_s $&$\frac{1}{\sqrt{3}}$\\\hline
\end{tabular}
\end{table}

\section{Non-Leptonic $\Xi_{bc}$ and $\Omega_{bc}$ decays}
\label{sec:bcq_nonleptonic}

The decays of $\Xi_{bc}$ and $\Omega_{bc}$ can proceed via the $b$ quark decay or the $c$ quark decay, which are induced by the following quark transitions:
\begin{eqnarray}
 c\to s \bar d u,  \;\;\; c\to u \bar dd/\bar ss, \;\;\; c\to  d \bar s u,\nonumber \\
 b\to \bar cc d/s, \;\; b\to c \bar ud/s, \;\; b\to u \bar cd/s, \;\; b\to q_1\bar q_2 q_3.
\end{eqnarray}

As we have shown in the semileptonic case, for the charm quark decays, one can obtain the decay amplitudes from those for $\Xi_{cc}$ and $\Omega_{cc}$ decays with the replacement of $T_{cc}\to T_{bc}$, $T_c\to T_b$ and $D\to B$.
For the bottom quark decay, one can obtain them  from those for $\Xi_{bb}$ and $\Omega_{bb}$ decays with $T_{bb}\to T_{bc}$, $T_b\to T_c$ and $B\to D$. Thus it is not necessary to repeat the tedious results here.

\section{Conclusions}
\label{sec:conclusions}

Quite recently, the LHCb collaboration has observed the doubly-charmed baryon $\Xi_{cc}^{++}$  in the   final state $\Lambda_c K^-\pi^+\pi^+$.  Such an important observation will undoubtedly promote the research on the hadron spectroscopy and also on   weak decays of   doubly  heavy baryons.

In this paper, we have analyzed    the weak decays of doubly heavy baryons
$\Xi_{cc}$,  $\Omega_{cc}$, $\Xi_{bc}^{(\prime)}$, $\Omega_{bc}^{(\prime)}$, $\Xi_{bb}$ and $\Omega_{bb}$ under   the flavor SU(3) symmetry.  Decay amplitudes for various semileptonic and nonleptonic decays have been parametrized in terms of a few  SU(3) irreducible amplitudes.  We have found a number of relations or sum rules between   decay widths and CP asymmetries, which can be examined in future measurements at experimental facilities like LHC, Belle II and CEPC. Moreover once a few decay branching fractions were measured in future,  some of these relations may provide hints for exploration  of new decay modes.

\section*{Acknowledgements}

The authors are   grateful to   Jibo He, Xiao-Hui Hu,     Cai-Dian L\"u,  Fu-Sheng Yu,  Zhen-Xing Zhao  for useful discussions and valuable comments. We also thank Prof. Xiao-Gang He and Prof. De-Shan Yang for discussions of the SU(3) decomposition.
This work is supported  in part   by National  Natural
Science Foundation of China under Grant
 No.11575110, 11655002, 11735010,  Natural  Science Foundation of Shanghai under Grant  No.~15DZ2272100 and No.~15ZR1423100, Shanghai Key Laboratory for Particle Physics and Cosmology, and  by    Key Laboratory for Particle Physics, Astrophysics and Cosmology, Ministry of Education.

\end{document}